\documentclass[iop]{emulateapj}
\usepackage{natbib}
\usepackage{epsfig}

\newcommand{\acounits}{\mbox{M$_\odot$ pc$^{-2}$ (K km s$^{-1}$)$^{-1}$}}

\shorttitle{CO Line Wings in Star-Forming ULIRGs}
\shortauthors{Leroy et al.}

\begin{document}

\slugcomment{Accepted for Publication in the Astrophysical Journal} 
\title{Faint CO Line Wings in Four Star-Forming (Ultra)luminous Infrared Galaxies}

\author{Adam K. Leroy\altaffilmark{1},
  Fabian Walter\altaffilmark{2},
  Roberto Decarli\altaffilmark{2},
  Alberto Bolatto\altaffilmark{3},
  Laura Zschaechner\altaffilmark{2},
  Axel Wei{\ss}\altaffilmark{4}}
\altaffiltext{1}{Department of Astronomy, The Ohio State University, 140 West 18th Avenue, Columbus, OH 43210}
\altaffiltext{2}{Max Planck Institute f\"ur Astronomie, K\"onigstuhl 17, 69117, Heidelberg, Germany}
\altaffiltext{3}{Department of Astronomy, University of Maryland, College Park, MD, USA}
\altaffiltext{4}{Max-Planck-Institute f\"ur Radioastronomie, Auf dem H\"ugel 69, D-53121 Bonn, Germany}

\begin{abstract}
We report the results of a search for large velocity width, low-intensity line wings --- a commonly used signature of molecular outflows --- in four low redshift (ultra)luminous infrared galaxies (U/LIRGs) that appear to be dominated by star formation. The targets were drawn from a sample of fourteen such galaxies presented in Chung et al. (2011), who showed the stacked CO spectrum of the sample to exhibit 1000~km~s$^{-1}$-wide line wings. We obtained sensitive, wide bandwidth imaging of our targets using the IRAM Plateau de Bure Interferometer. We detect each target at very high significance but do not find the claimed line wings in these four targets. Instead, we constrain the flux in the line wings to be only a few percent. Casting our results as mass outflow rates following Cicone et al. (2014) we show them to be consistent with a picture in which very high mass loading factors preferentially occur in systems with high AGN contributions to their bolometric luminosity. We identify one of our targets, IRAS05083 (VII Zw 31), as a candidate molecular outflow.
\end{abstract}


\section{Introduction}
\label{sec:intro}

Galactic winds \citep{VEILLEUX05} can carry gas, metals, and dust from a galaxy to its halo, deplete the future fuel for star formation, and inject energy and momentum into the inter- and circum-galactic medium. Winds seen in neutral gas via absorption lines are  a near-ubiquitous feature of star-forming galaxies \citep{RUPKE05}. In galaxies with strong active galactic nuclei (AGN), there is also evidence for widespread molecular winds \citep[e.g.,][]{FERUGLIO10,ALATALO11,VEILLEUX13,CICONE14,CICONE15} with suggestions of a correlation between AGN strength and the amount of outflowing material \citep{CICONE14}. There is also evidence for molecular outflows, or at least material blown out of the disk, in galaxies dominated by star formation \citep[e.g.,][]{WALTER02,BOLATTO13A,SAKAMOTO14}. However, the number of very powerful star formation-driven molecular winds remains substantially lower than the set of molecular winds believed to be driven by AGN.

An excellent prospect to build a large sample of star formation-driven molecular winds comes from a result by \citet[][hereafter C11]{CHUNG11} based on observations of ultraluminous infrared galaxies (ULIRGs) by \citet{CHUNG09}. \citet{CHUNG09} observed CO emission from 29 ULIRGs using the wide-bandwidth Redshift Search Receiver instrument, at the time mounted on the 14-m Five Colleges Radio Astronomy Observatory. C11 stacked this sample into a single high signal-to-noise spectrum and noted that this spectrum showed very wide ($\approx 1000$~km~s$^{-1}$), line wings that accounted for, on average, $\sim 25\%$ of the total CO emission from the galaxy. This is exactly the sort of signal expected from a powerful outflow of molecular gas from a moderately inclined galaxy. It resembles the signatures seen in individual sources like NGC 1266 \citep{ALATALO11}, Mrk 231 \citep{FERUGLIO10,MAIOLINO12}, and the \citet{CICONE14} sample of galaxies. The result of C11 is especially interesting because they showed that the stacked signal appeared to come preferentially from galaxies whose spectroscopic classification was either ``Starburst'' or ``\ion{H}{2} region'' rather than Seyfert or LINER. Thus, the C11 sample offers one of the best prospects for building a sample of powerful molecular winds driven primarily by star formation.

\section{Observations}
\label{sec:obs}

Based on the claim by C11, during the summer of 2014 we used the IRAM Plateau de Bure Interferometer\footnote{This work is based on observations carried out with the IRAM Plateau de Bure Interferometer. IRAM is supported by INSU/CNRS (France), MPG (Germany), and IGN (Spain)} to observe four members of the C11 Starburst/\ion{H}{2} subsample. We targeted IRAS05083+7936 (I05083), IRAS10035+4852 (I100035), IRAS17132+5313 (I17132), and IRAS17208-0014 (I17208), which C11 classified as bright (I05083, I17208) or intermediate (I100035, I17132) in CO flux. Table \ref{tab:source} lists these targets and their basic physical parameters. We observed each target in the PdBI's most compact configuration for one track ($\approx 4$ hours on source), targeting the $^{12}$CO (1-0) line using the WideX correlator in its maximum bandwidth configuration. The resulting data have synthesized beams of $5.0 \times 4.7\arcsec$ (I05083), $6.4 \times 4.4\arcsec$ (I10035), $4.7 \times 3.8\arcsec$ (I17132), and $5.6 \times 4.4\arcsec$ (I17208) and noise of 0.9 (I05083), 1.4 (I10035), 0.9 (I17132), and 1.5 (I17132)~mJy~beam$^{-1}$. The data were imaged and deconvolved with a channel width of 25~km~s$^{-1}$ following standard procedures in GILDAS. Before any other analysis, we subtracted the continuum in the image plane using line free channels to either side of the CO line and not including the region of the CN lines, which are the only other bright lines in our bandpass. The peak 3mm continuum flux density in our targets is $\approx 0.8 \pm 0.04$~mJy~beam$^{-1}$ (I05083), $< 0.36$~mJy~beam$^{-1}$ (5$\sigma$, I10035), and $0.63 \pm 0.03$~mJy~beam$^{-1}$ (17132), $6.3 \pm 0.17$~mJy~beam$^{-1}$ (I17208). For I05083 and I1713, these are consistent with a synchrotron spectral index and the fluxes of a few mJy measured at 33~GHz \citep{LEROY11B}. For I17208, the 6~mJy flux is slightly lower than the $\approx 9 \pm 1$~mJy measured at 85~GHz by \citet{IMANISHI06}, but consistent within likely calibration uncertainties, especially if the emission has a synchrotron spectral index.

A few residual sidelobes remain present in channels containing the main CO line for the two brightest targets (I05083 and I17208; see the peak intensity maps in Figure \ref{fig:maps}). These could be removed with more aggressive cleaning but do not matter to the results of this paper. The bright line is already almost entirely cleaned and we search for line wings in channels away from the main line, where these sidelobes are not present. The dynamic range in the images is still very high, 350 in the highest case (I05083).

In this most compact configuration, the PdBI is expected to resolve out sources larger than about 15$\arcsec$ in an individual channel. This is much larger than the visible extent of any of our targets and the corresponding physical distance ($\approx 15$~kpc) is larger than the extent of any known molecular outflows at a fixed velocity \citep{CICONE14}.  In our observations we find 96, 93, 82, and 130\% of the flux in the C11 FCRAO spectrum and the spectral shapes match well for the bright part of the line (see below). Given this good match to the single dish data and the large distance to our sources (so that even large physical scales are compact on the sky; $15\arcsec \approx 15$~kpc), we do not expect spatial filtering by the interferometer to be an important consideration for this work.

To calculate CO luminosities and sizes, we follow \citet{WRIGHT06} using standard parameters ($H_0 = 70$~km~s$^{-1}$~Mpc$^{-1}$, $\Omega_M = 0.3$, $\Omega_\Lambda = 0.7$; consistent with the references used for luminosity). For these parameters and a typical $z = 0.05$ the luminosity distance is $222$~Mpc and angular size distance is $202$~Mpc, so that $1\arcsec \approx 1$~kpc.

\begin{deluxetable}{lcccc}
\tablecaption{Observed targets} 
\tablehead{ 
\colhead{Galaxy} & 
\colhead{$\log_{10} L_{TIR}$} & 
\colhead{$f_{\rm AGN}$} & 
\colhead{$SFR$} &
\colhead{$\left< z_{\rm CO} \right>$} 
\\
\colhead{} & 
\colhead{(log$_{10}$ L$_\odot$)} &
\colhead{} &
\colhead{M$_\odot$~yr$^{-1}$} &
\colhead{}
}
\startdata
IRAS05083+7936 & 11.9 & $< 0.1^\tablenotemark{a}$ & $\sim 90$ & 0.054 \\
IRAS10035+4852 & 12.0 & $< 0.33$ & $\sim 100$ & 0.065 \\
IRAS17132+5313 & 11.9 & $\sim 0.17$ & $\sim 60$ & 0.051 \\
IRAS17208-0014 & 12.4 & $< 0.1$ & $\sim 250$ & 0.043 
\enddata
\label{tab:source}
\tablenotetext{a}{Adopted based on high 6.2$\mu$m equivalent width
of 0.64 (highly star forming) from \citet{STIERWALT13}. For comparison, I17208 has 0.31.}
\tablecomments{IR luminosities from \citet{SANDERS03} except I10035 from \citet{HWANG07}. 
AGN fraction estimates from \citet{FARRAH03}, see also \citet{KOSS13} for a hard $X$-Ray upper limit on I17208. SFR translated from IR following \citet{CICONE14} for consistency.}
\end{deluxetable}

\section{Results}
\label{sec:results}

\begin{figure*}
\includegraphics[width=1.75in]{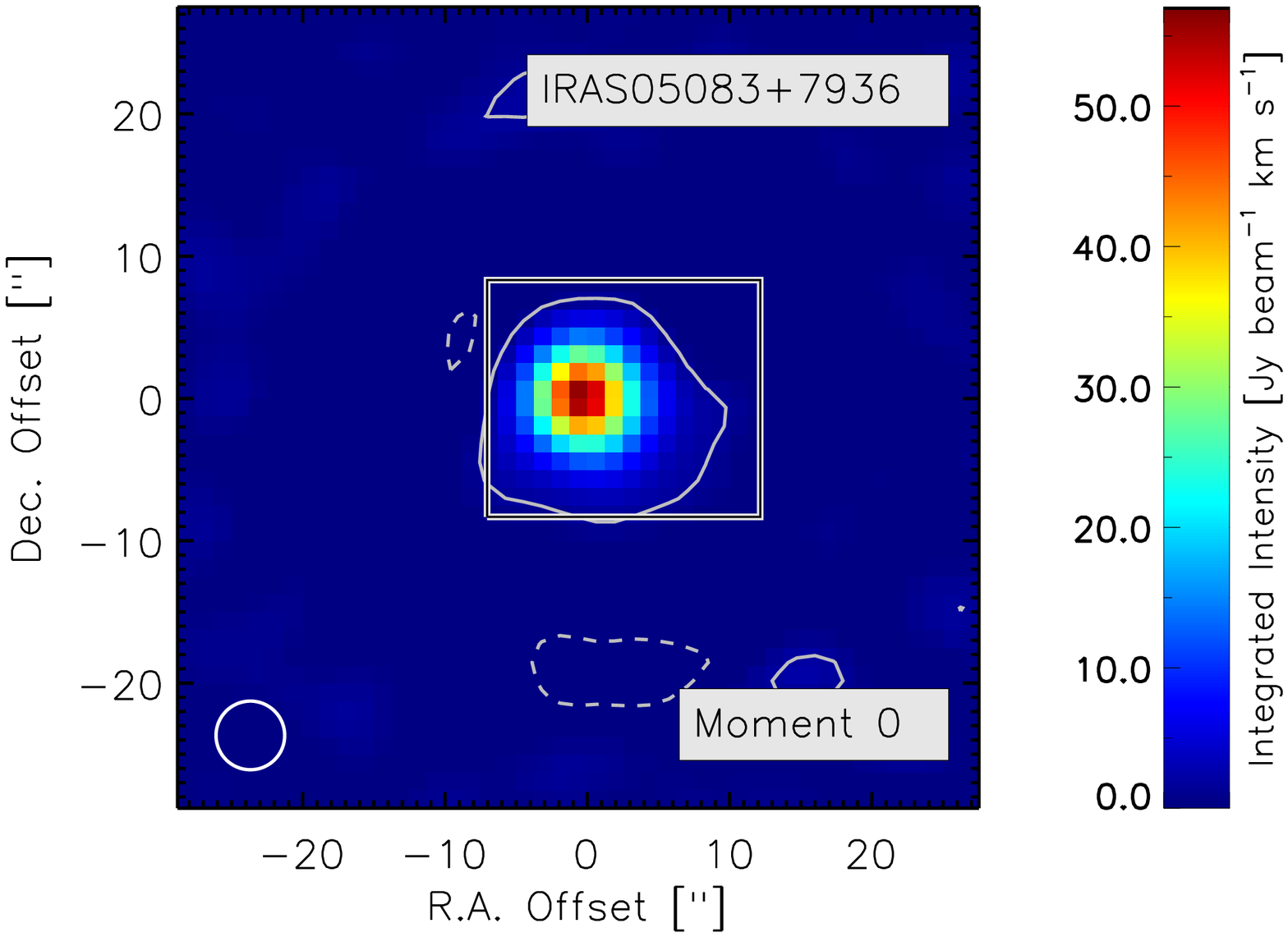}
\includegraphics[width=1.75in]{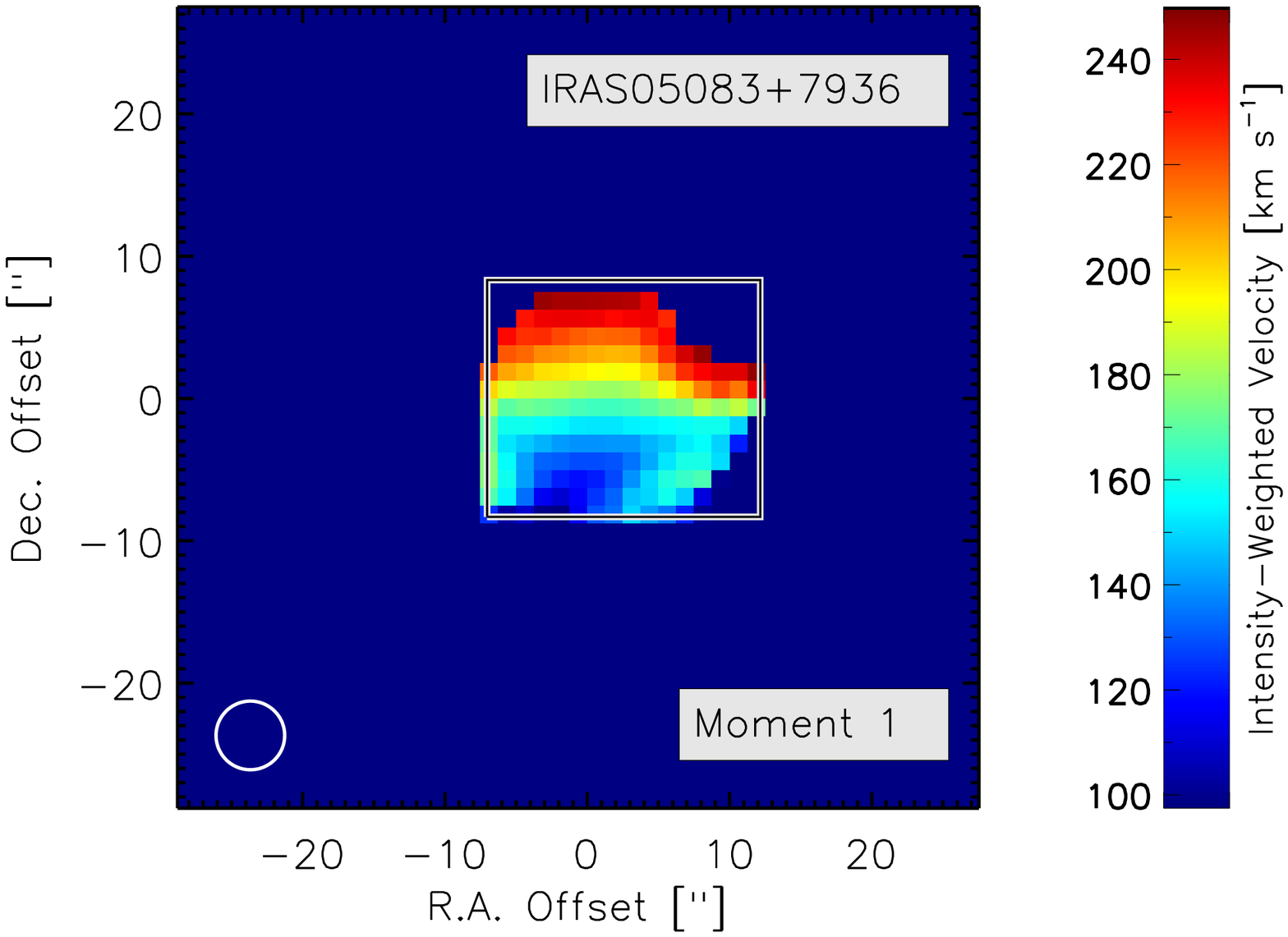}
\includegraphics[width=1.75in]{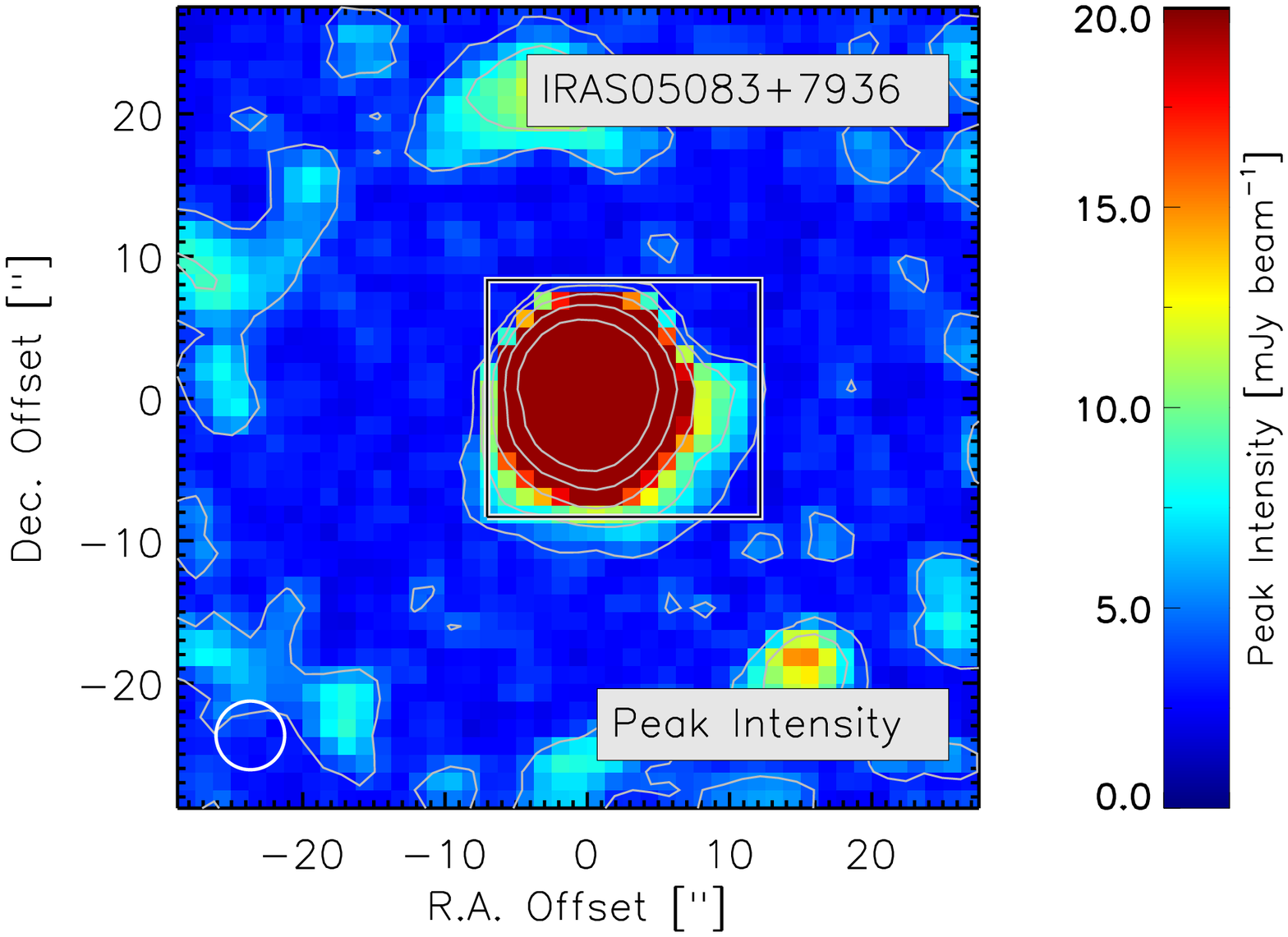}
\includegraphics[width=1.75in]{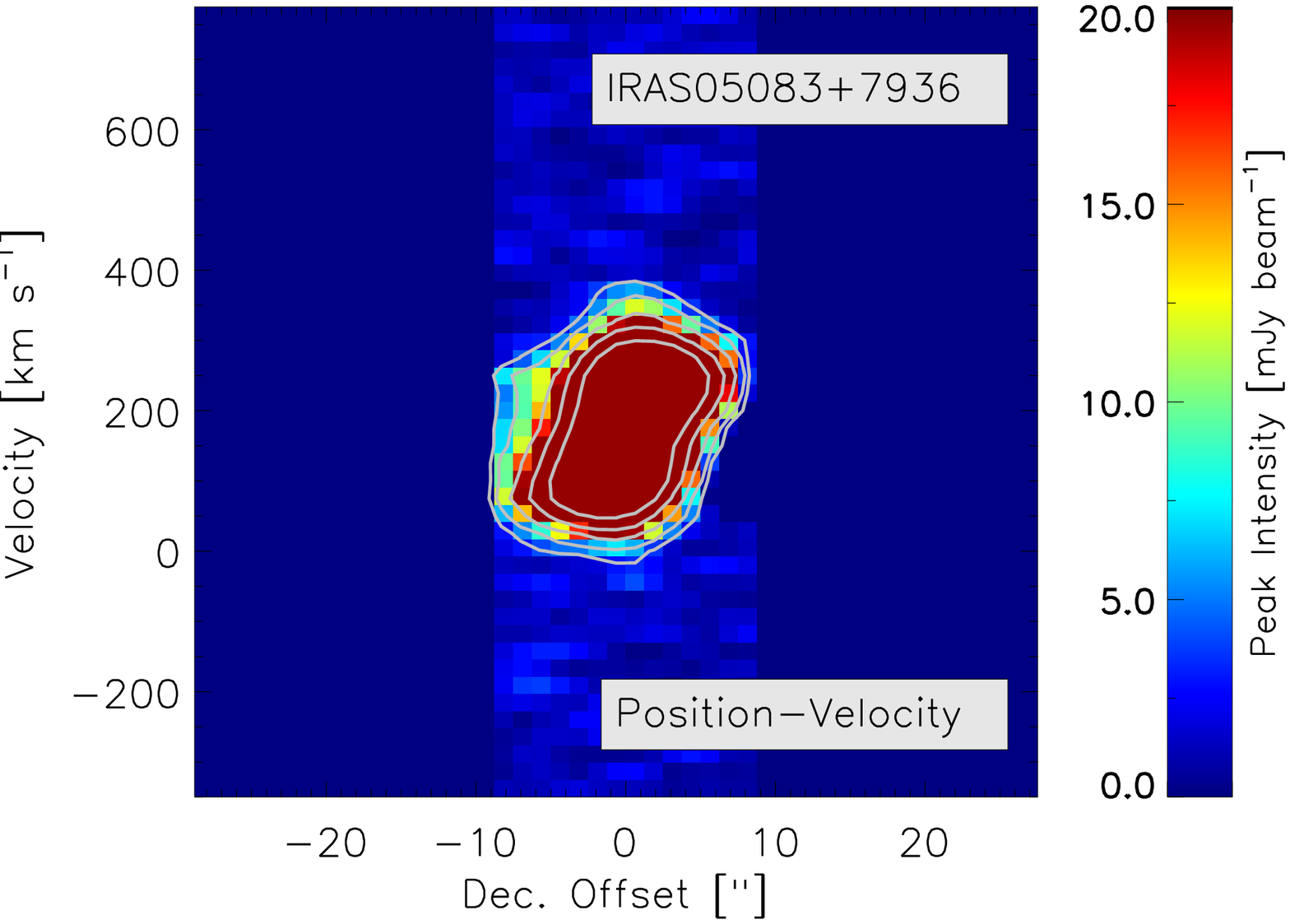}
\includegraphics[width=1.75in]{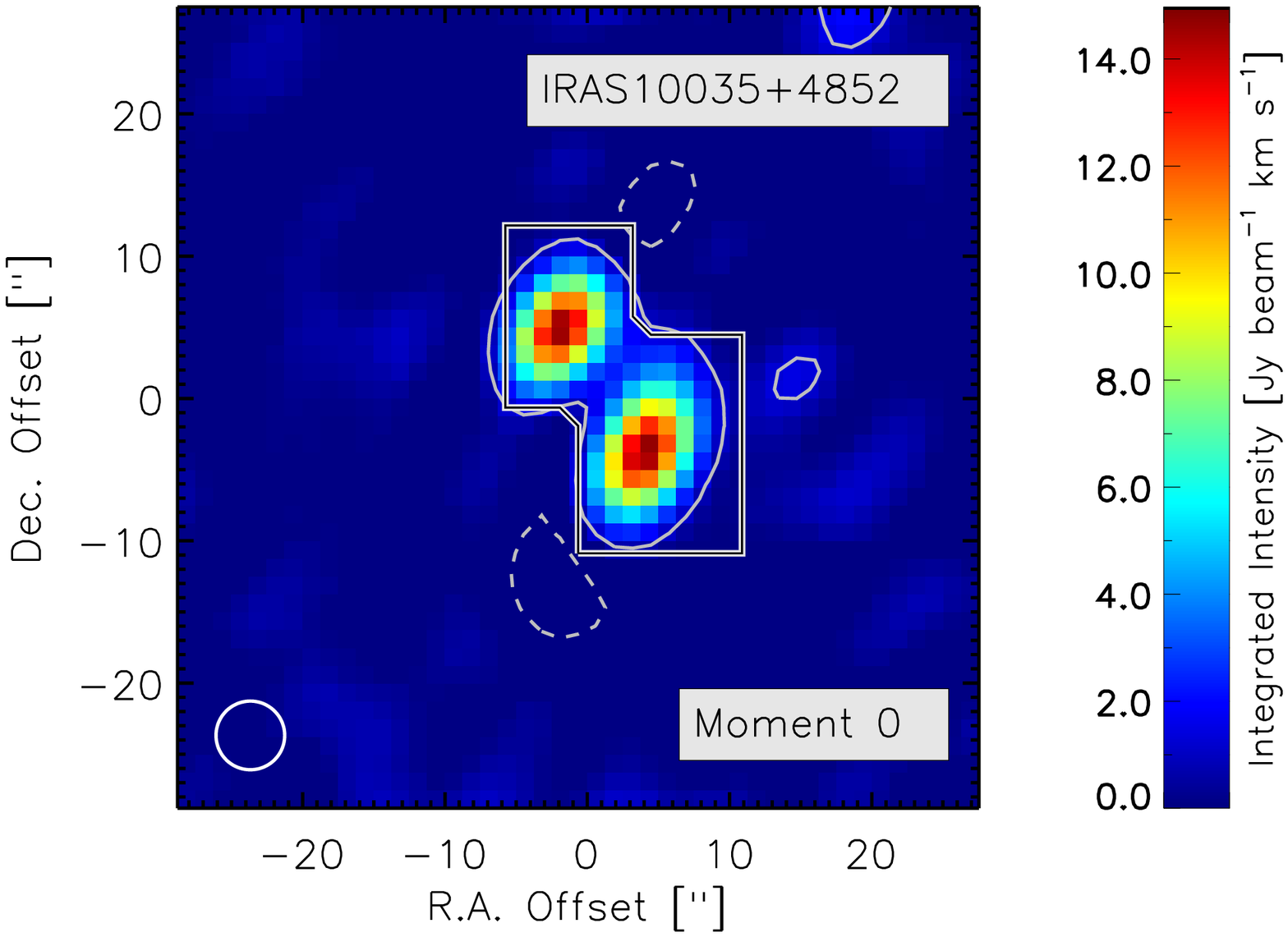}
\includegraphics[width=1.75in]{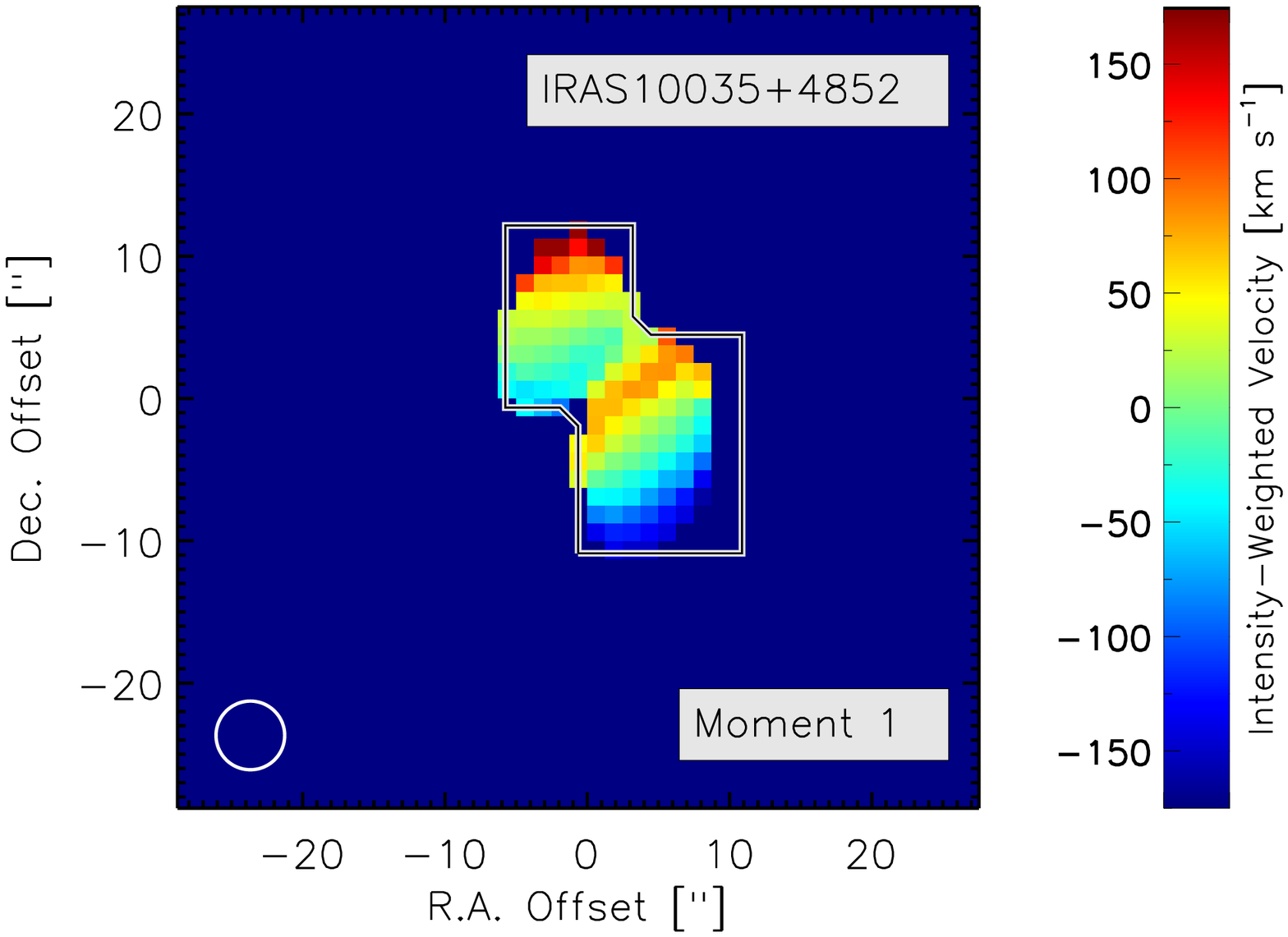}
\includegraphics[width=1.75in]{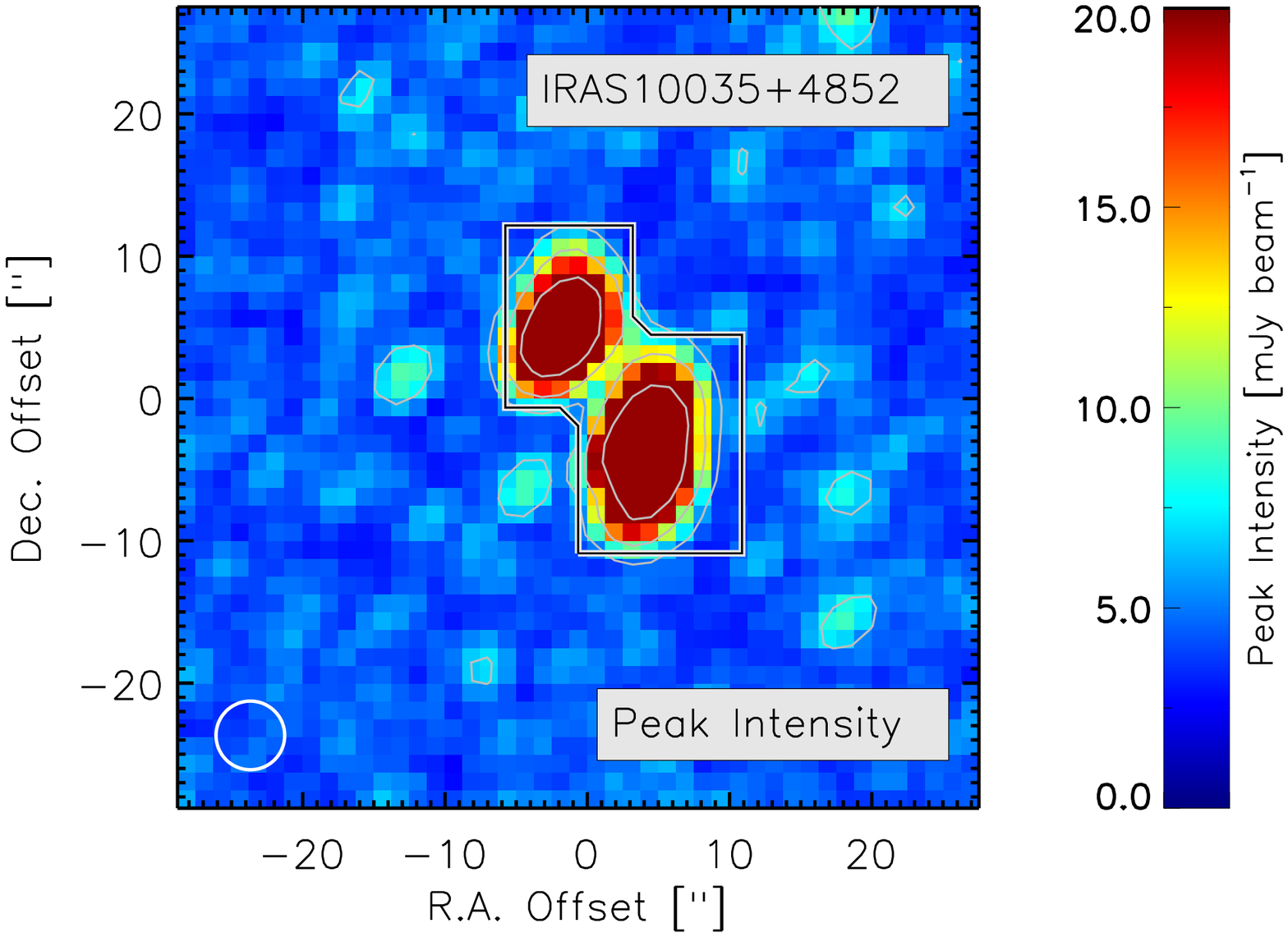}
\includegraphics[width=1.75in]{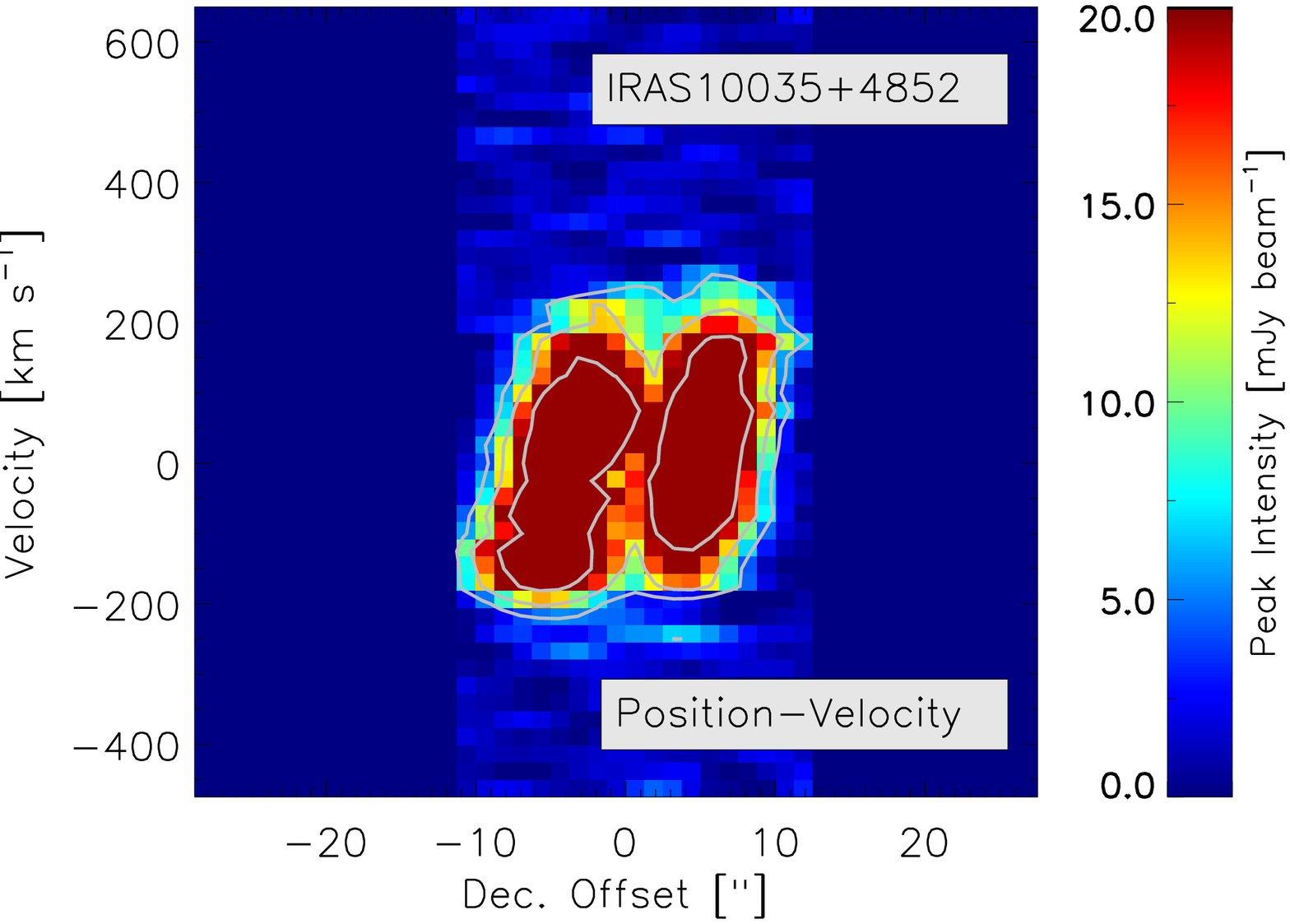}
\includegraphics[width=1.75in]{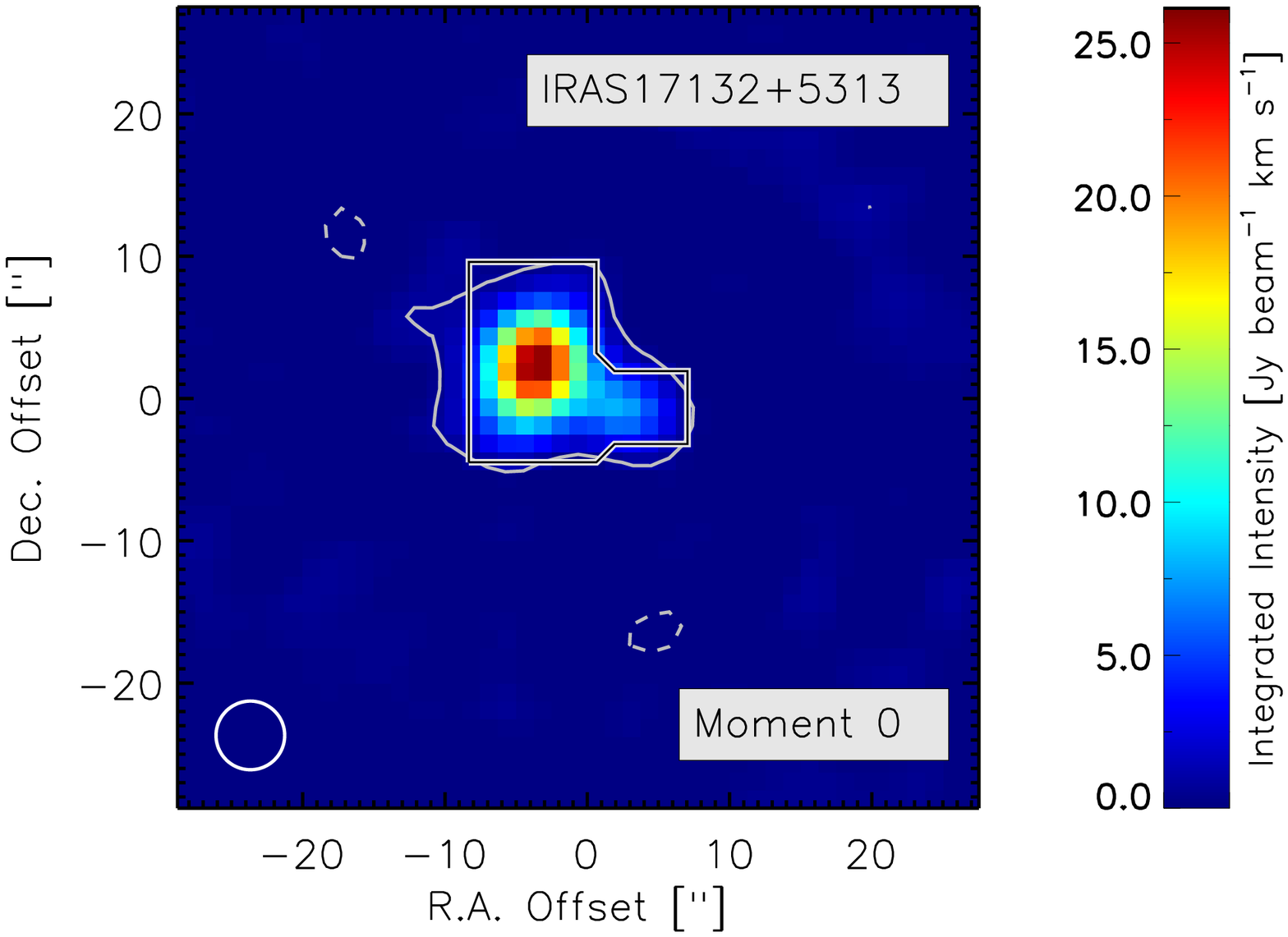}
\includegraphics[width=1.75in]{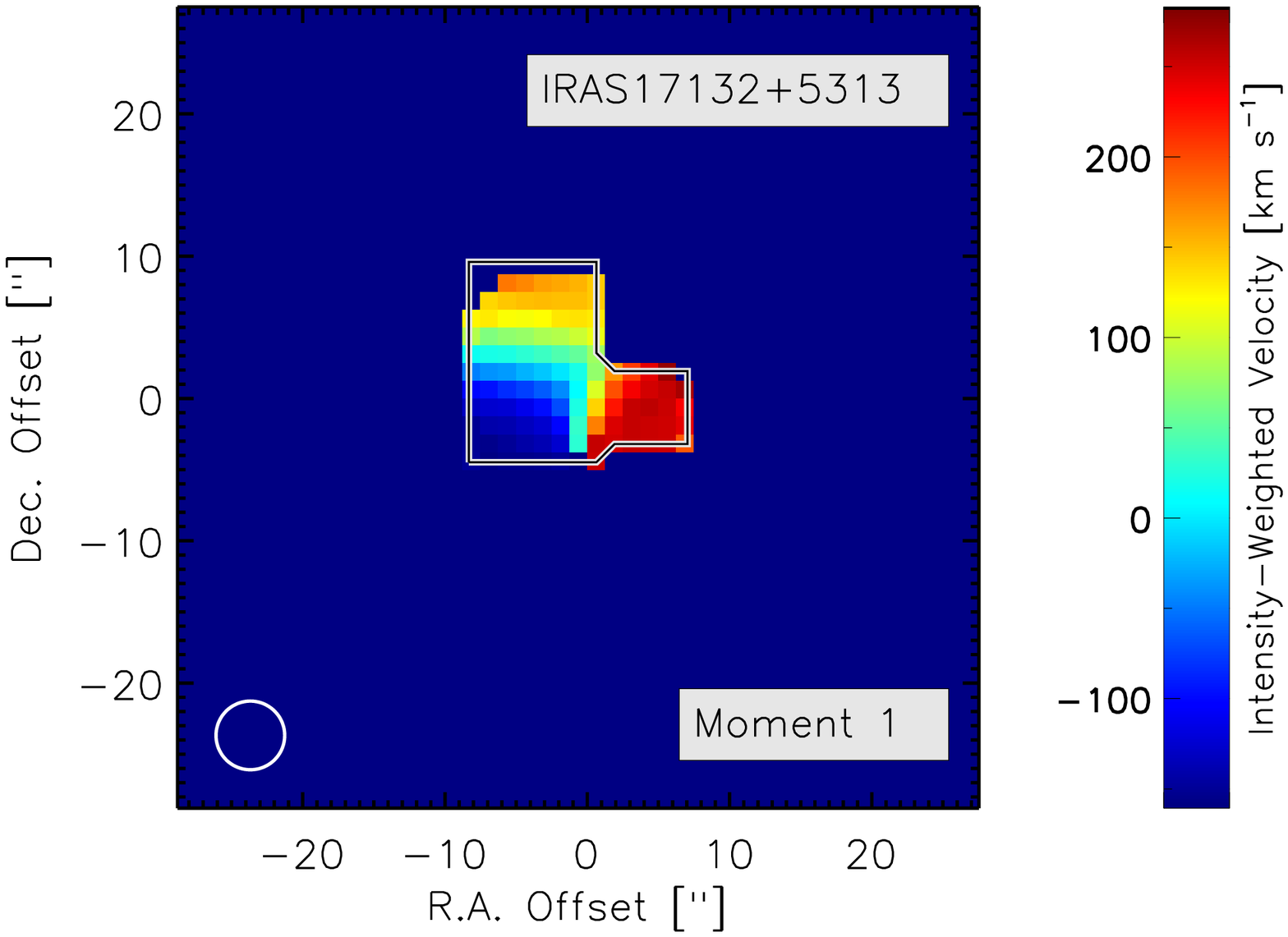}
\includegraphics[width=1.75in]{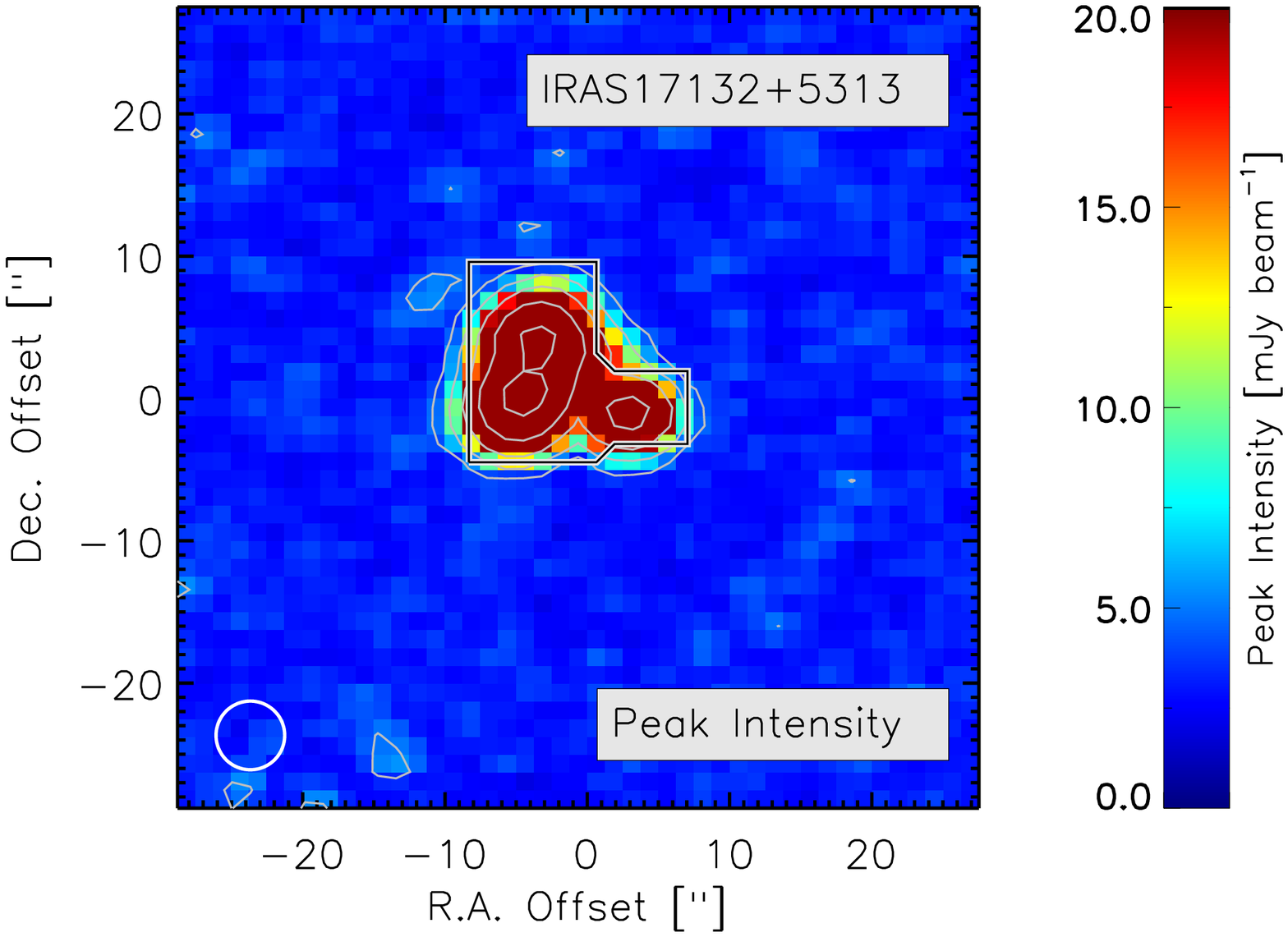}
\includegraphics[width=1.75in]{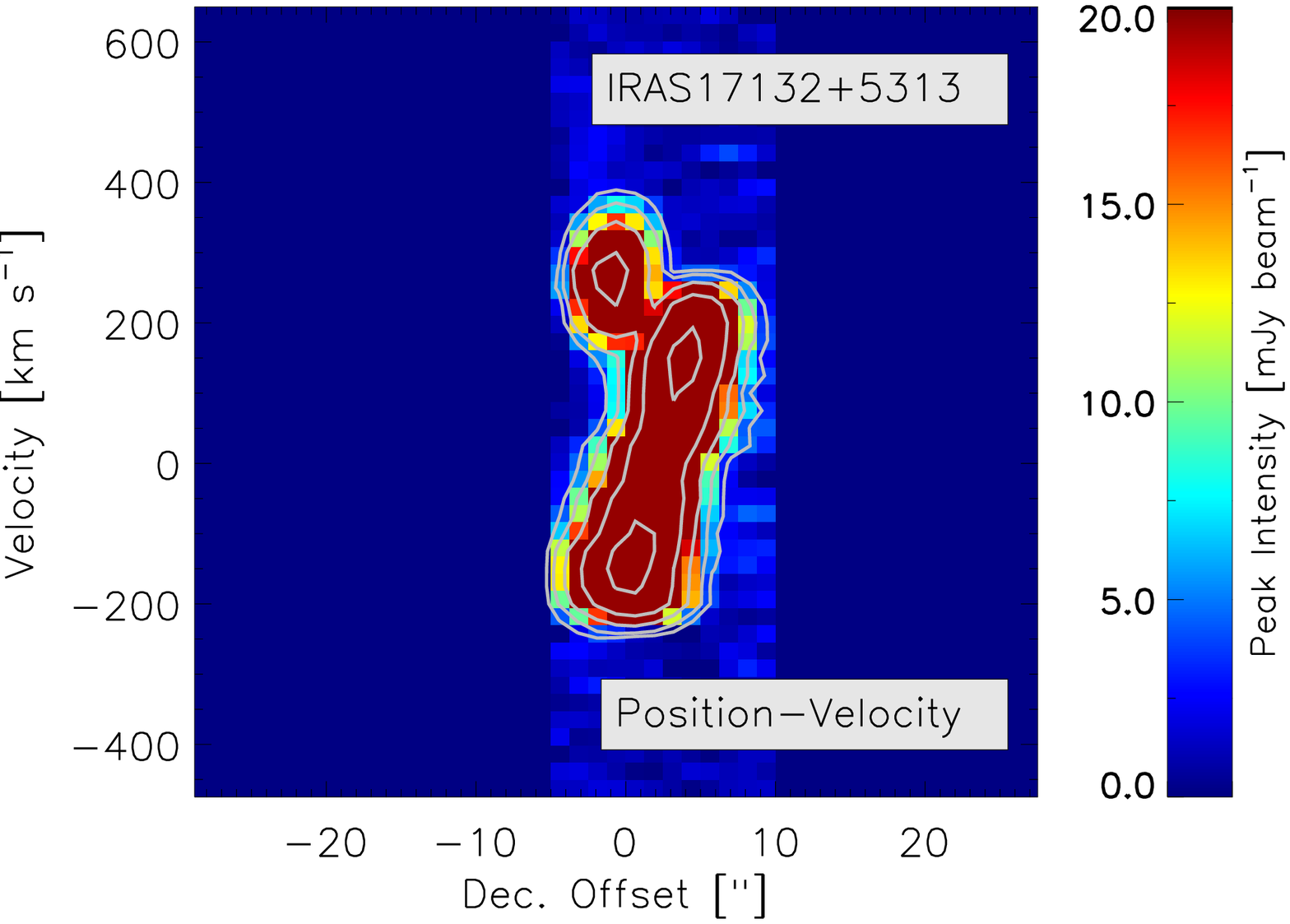}
\includegraphics[width=1.75in]{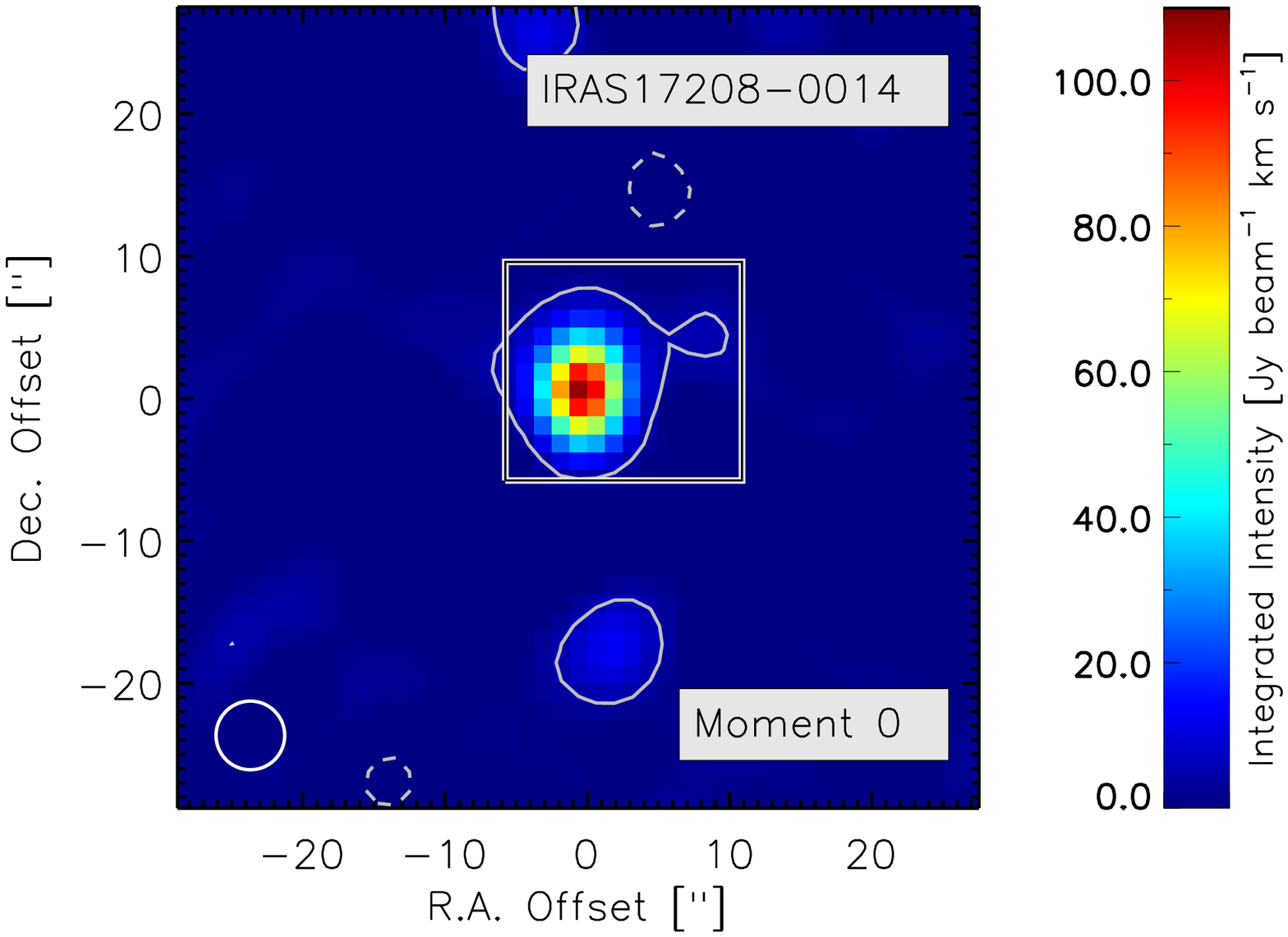}
\includegraphics[width=1.75in]{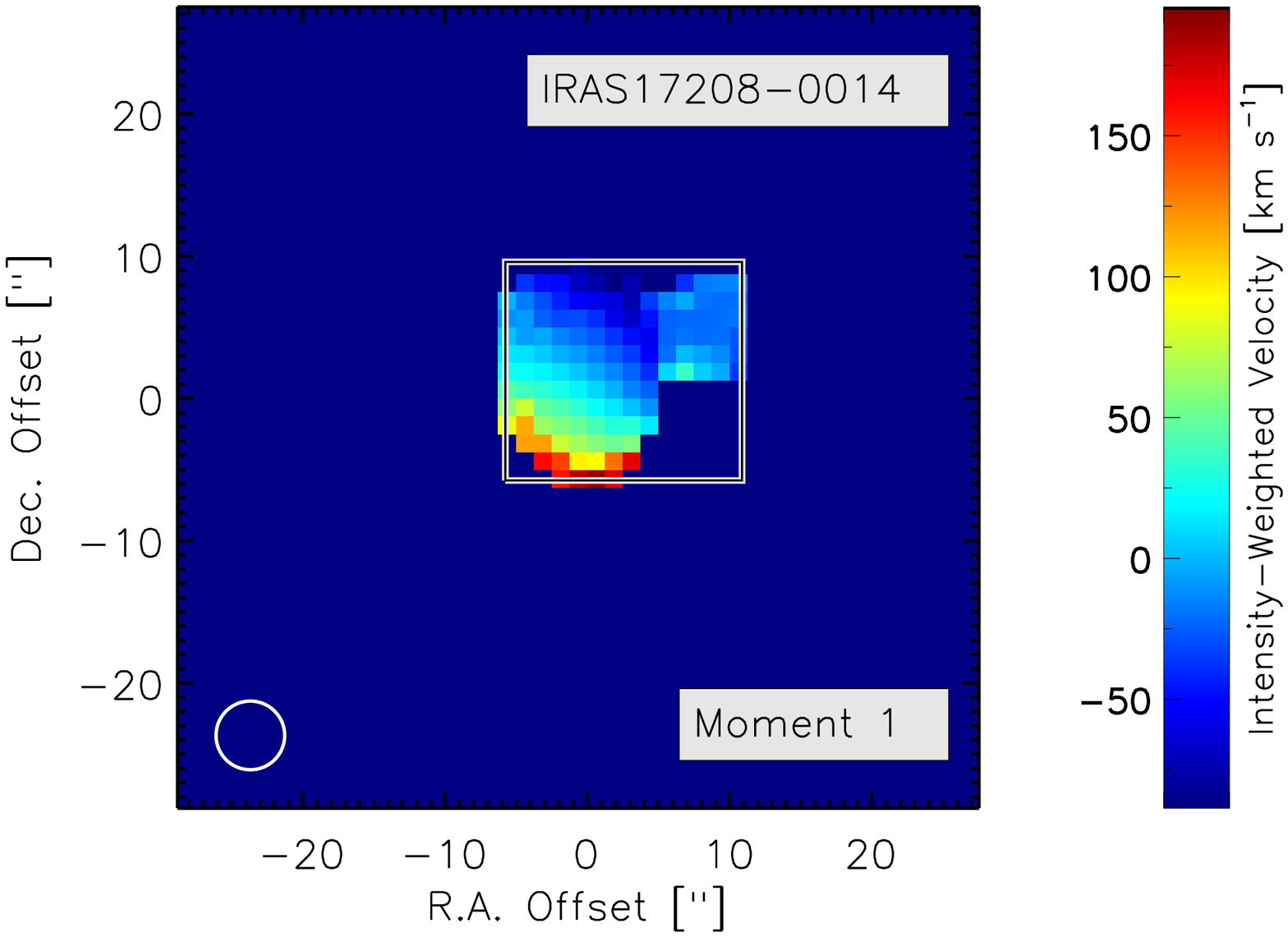}
\includegraphics[width=1.75in]{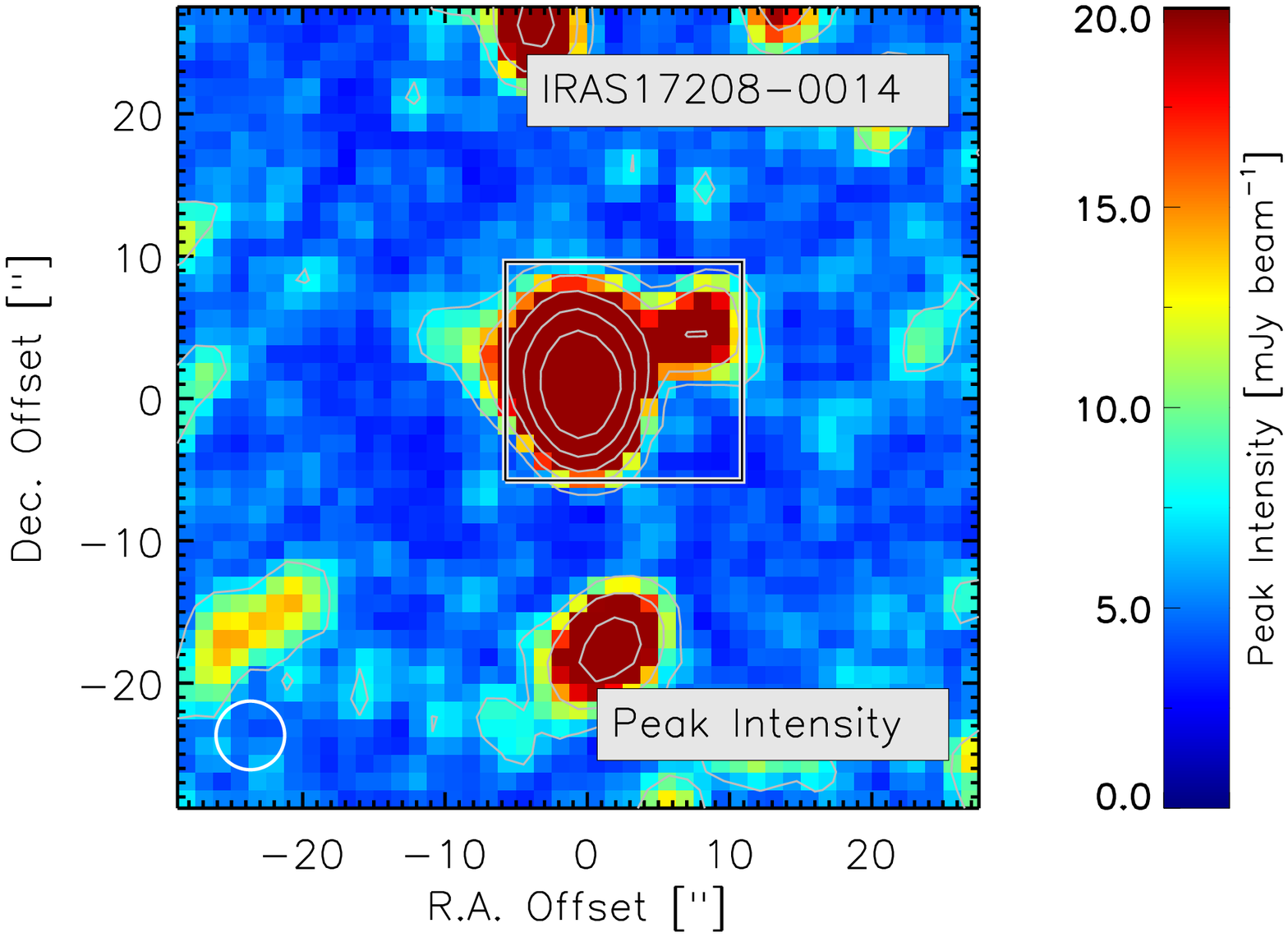}
\includegraphics[width=1.75in]{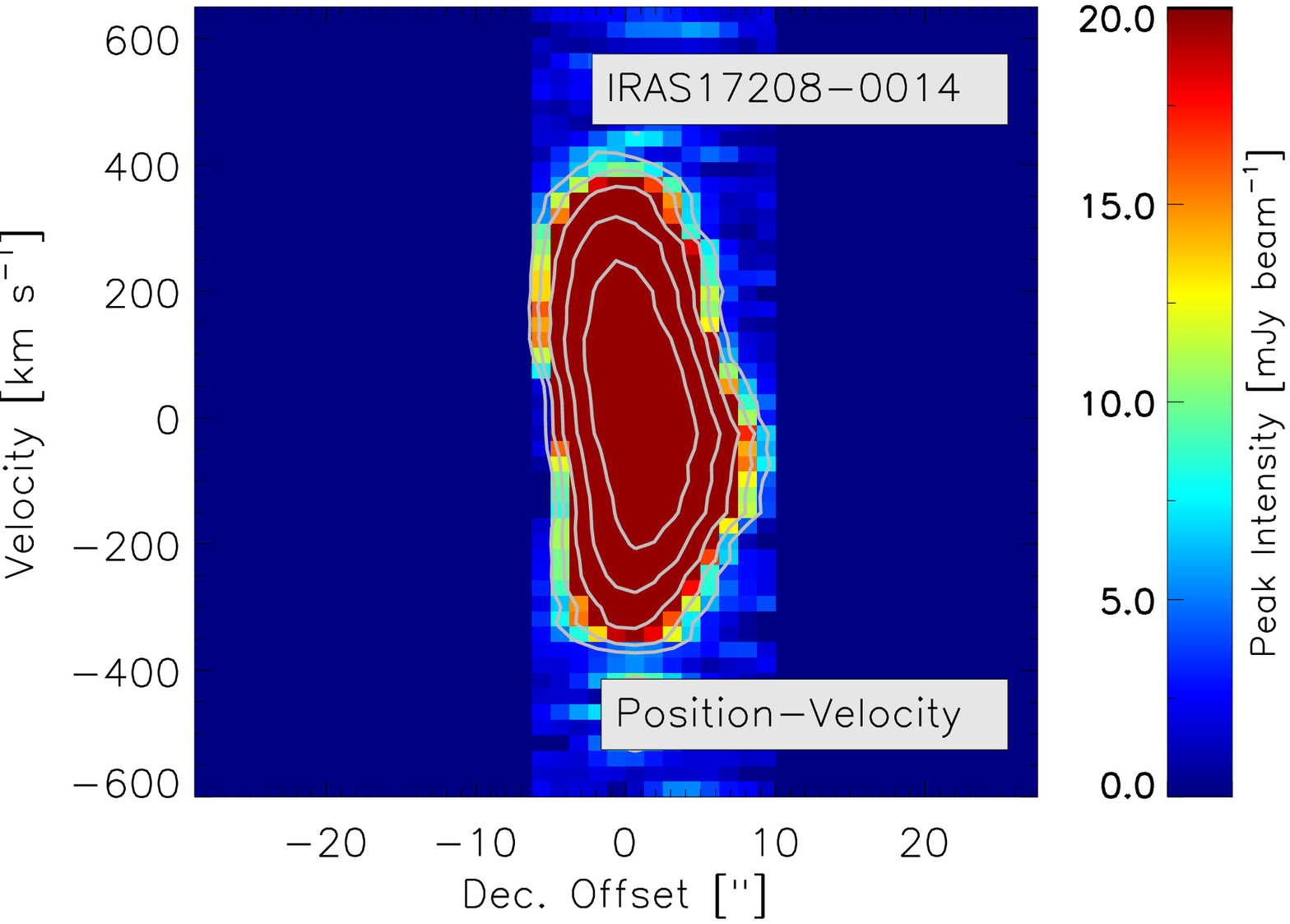}
\caption{CO observations of our targets. ({\em far left}) Integrated intensity maps. Gray contours mark regions with SNR $>3$ (solid) and $<-3$ (dashed). Boxes in the first three panels show regions of interest used to construct the spectra used below. ({\em middle left}) Intensity weighted velocity field constructed from emission with SNR$>5$. Velocity gradients, possibly indicative of rotation, are visible in all four targets. ({\em middle right}) Peak intensity maps for each target, saturated at 20~mJy~beam$^{-1}$ with contours at 3$\sigma$ and increasing by factors of $2$. ({\em far right}) Peak intensity measured along the declination axis (which aligns roughly with the velocity gradient) within the region of interest. A circle indicates the beam extent in the left two images. Coordinates are plotted in each panel as arcsecond offsets from the pointing center; typically $1\arcsec \approx 1$~kpc. Offset $(0, 0)$ corresponds to the pointing center. I10035 and I17132 are each resolved into two discrete components. The offset structures around I05083 and I17208 are residual sidelobes in the main bright part of the line and not relevant to the current analysis.
\label{fig:maps}}
\end{figure*}

\begin{deluxetable*}{lcccccc}
\tablecaption{CO Properties From Integrated Spectra} 
\tablehead{ 
\colhead{Galaxy} & 
\colhead{Noise\tablenotemark{a}} &
\colhead{$w_{20}^{\rm CO}$\tablenotemark{b}} & 
\colhead{$F_{\rm CO}$} & 
\colhead{$L_{\rm CO}$} &
\colhead{$F^{\rm wings}_{\rm CO}$\tablenotemark{c}} & 
\colhead{$f^{\rm wings}_{\rm CO}$\tablenotemark{d}} 
\\
\colhead{} & 
\colhead{(mJy)} & 
\colhead{(km s$^{-1}$)} & 
\colhead{(Jy~km s$^{-1}$)} & 
\colhead{($10^9$ K~km~s$^{-1}$~pc$^2$)} & 
\colhead{(Jy~km s$^{-1}$)} &
\colhead{}
}
\startdata
IRAS05083 & 2.0 & $250$ & $100.2 \pm 0.4$ & $13.5 \pm 0.06$ & $1.1 \pm 0.5$ & $0.011 \pm 0.005$ \\
IRAS10035 & 2.4 & $400$ & $38.4 \pm 0.5$ & $7.5 \pm 0.1$ & $0.9 \pm 0.6$ & $< 0.045$ \\
IRAS17132 & 1.9 & $550$ & $54.3 \pm 0.4$ & $6.6 \pm 0.05$ & $0.1 \pm 0.5$ & $< 0.026$ \\
IRAS17208 & 4.3 & $550$ & $140.5 \pm 1.0$ & $12.2 \pm 0.08$ & $3.2 \pm 1.0$ & $0.026 \pm 0.007$ \enddata
\label{tab:obs}
\tablenotetext{a}{Noise in the integrated spectrum using 25~km~s$^{-1}$ channels.}
\tablenotetext{b}{Velocity width at 20\% of the peak value of the CO line.}
\tablenotetext{c}{Flux in a broad, low component obtained by integrating the spectrum where $|v| < 1500$~km$^{-1}$ and the flux is less than 5\% of the maximum value. This is the gray area in Figure \ref{fig:spectra}.} 
\tablenotetext{d}{The flux of the broad component expressed as a fraction of the total flux. For a Gaussian profile, one expects $f_{\rm wings} \approx 0.014$. Limits are 3$\sigma$ for IRAS10035 and IRAS17132.} 
\tablecomments{Statistical uncertainties are reported in the table, but the main uncertainty on $w_{20}$, $F_{\rm CO}$, and $L_{\rm CO}$ is systematic, not statistical. We adopt $\approx 15\%$ plus uncertainty in the Hubble flow for $F_{\rm CO}$ and $L_{\rm CO}$ and $\pm 25$~km~s$^{-1}$ for $w_{20}$.}
\end{deluxetable*}

We detected bright $^{12}$CO emission from all of our sources. For each target, Figure \ref{fig:maps} shows the integrated intensity (``moment 0'') maps (far left), intensity-weighted velocity (``moment 1'') maps (middle left), and maps of the peak intensity both along each line of sight as a function of sky position (middle right) and collapsed along the right ascension\footnote{We choose to measure the position-velocity diagram along the declination axis because our targets (conveniently) show velocity gradients mostly lined up in this direction.} axis (far right). The peak intensity maps saturate at 20~mJy~beam$^{-1}$ in order to show extended structure. In the first three panels, boxes show the region of interest encompassing bright emission for each target. We use this region to construct an integrated spectrum for each target.

In two sources, I100035 and I17132, we resolve the source into multiple components. In I10035 the two sources each show a velocity gradient that is likely rotation and so appear to be individual galaxies. In I17132, the second component is known to be a second, fainter galaxy from optical imaging \citep{ARMUS90}. I17208 also shows some faint extension to the northwest, but this appears approximately consistent with an extension of the velocity field of the rotating disk. This system does have two components \citep[see][]{GARCIABURILLO15} but they are not resolved in our observations. Therefore we treat this target as having only a single component.

Table \ref{tab:obs} reports the integrated properties of the CO line in each component. We integrate fluxes over the range $\pm 1000$~km~s$^{-1}$ around the nominal redshift, report the width of CO emission at 20\% of its peak value (indicated by the dashed lines in Figure \ref{fig:spectra}), and convert the line flux to a luminosity using the distances described above. The line widths are reasonable for massive disks, $200$--$600$~km~s$^{-1}$, and comparable to other ULIRGS \citep[e.g., see][]{CHUNG09}. Gradients, which are most naturally interpreted as signatures of rotation, are clearly  visible in the velocity field maps, though with $\approx 5$~kpc resolution, we only marginally resolve each target. The luminosities are typical of ULIRGs and consistent with previous measurements (see above).

\subsection{Search for Faint Line Wings}

\begin{figure*}
\plottwo{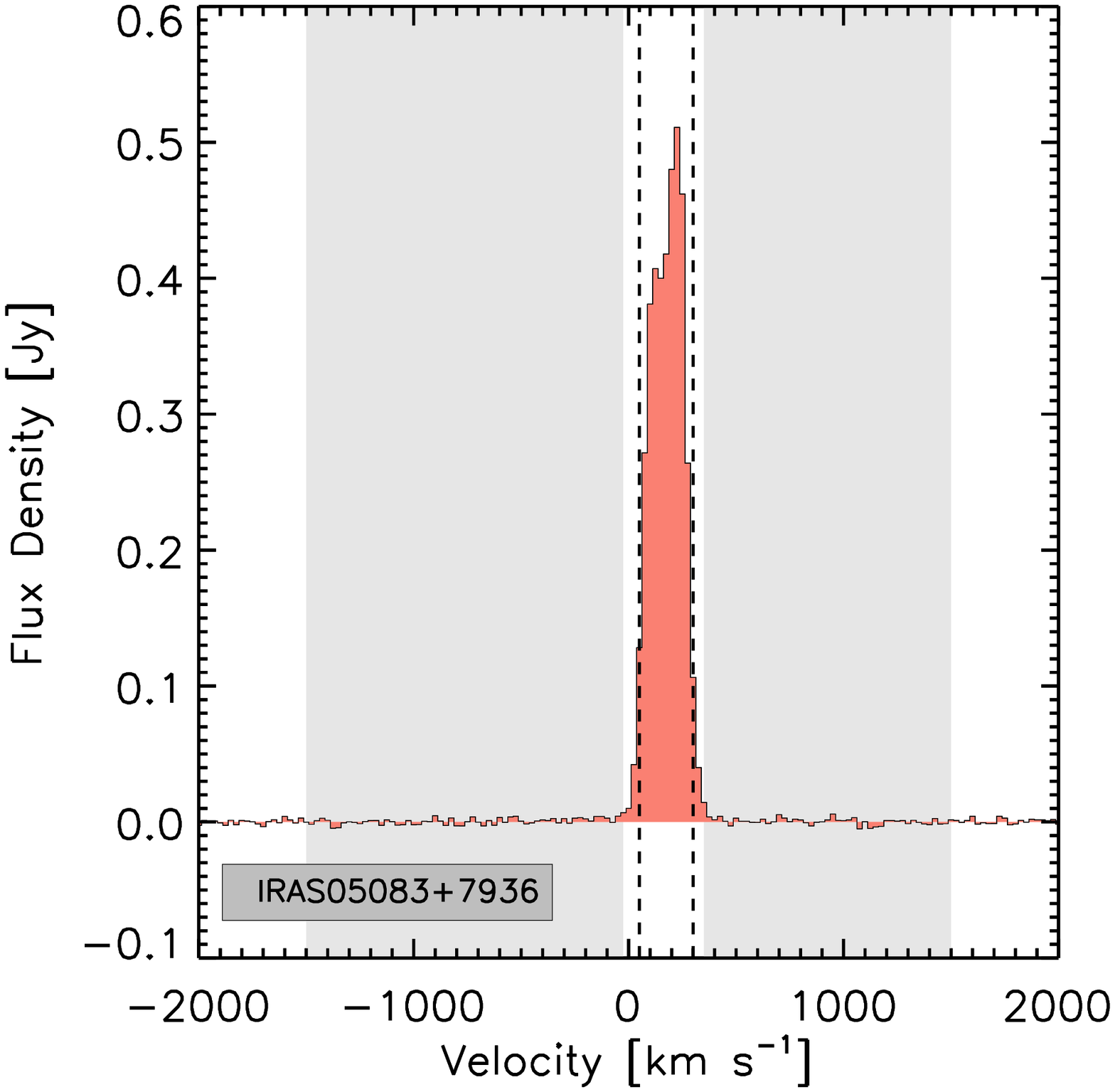}{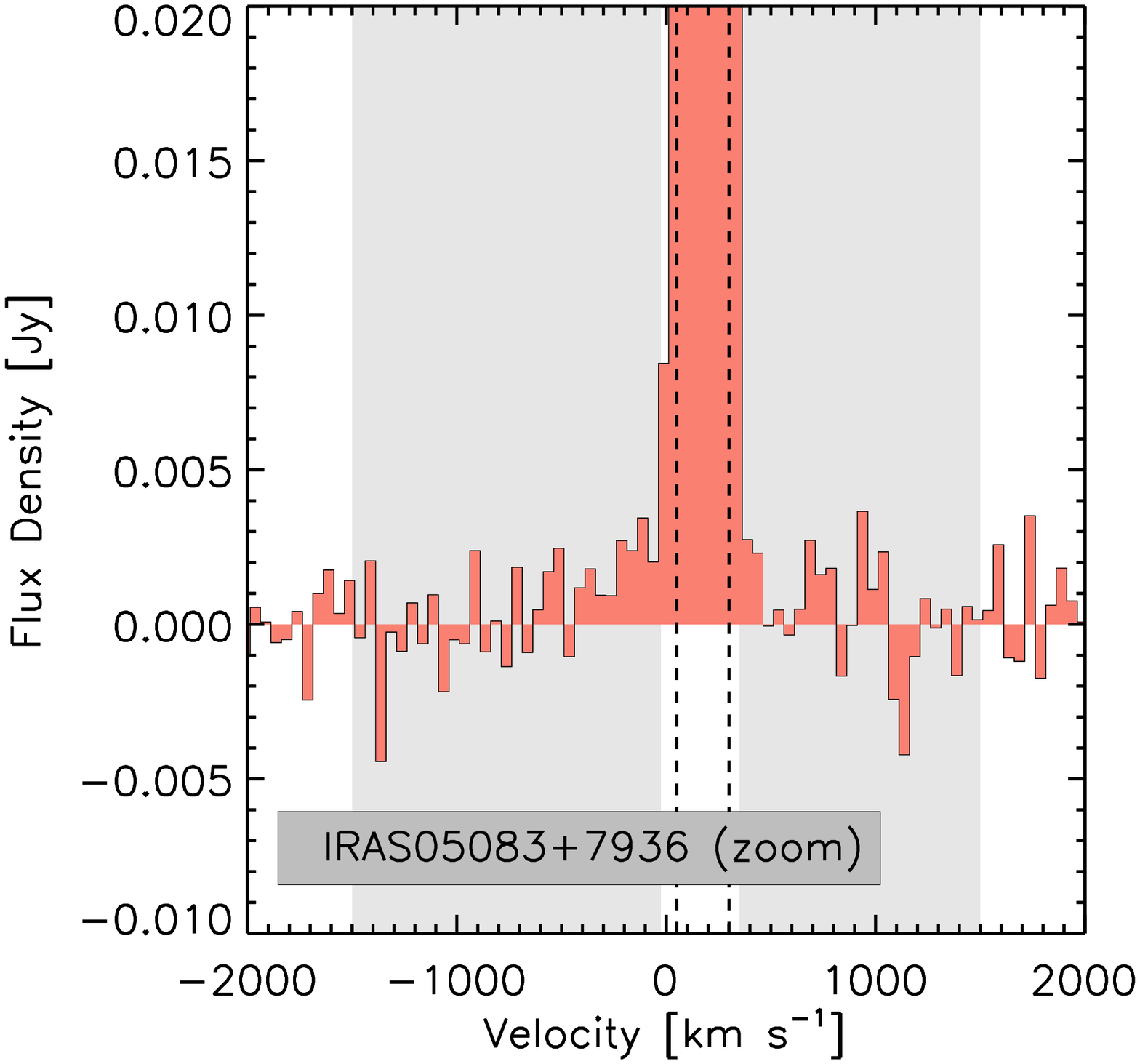}
\plottwo{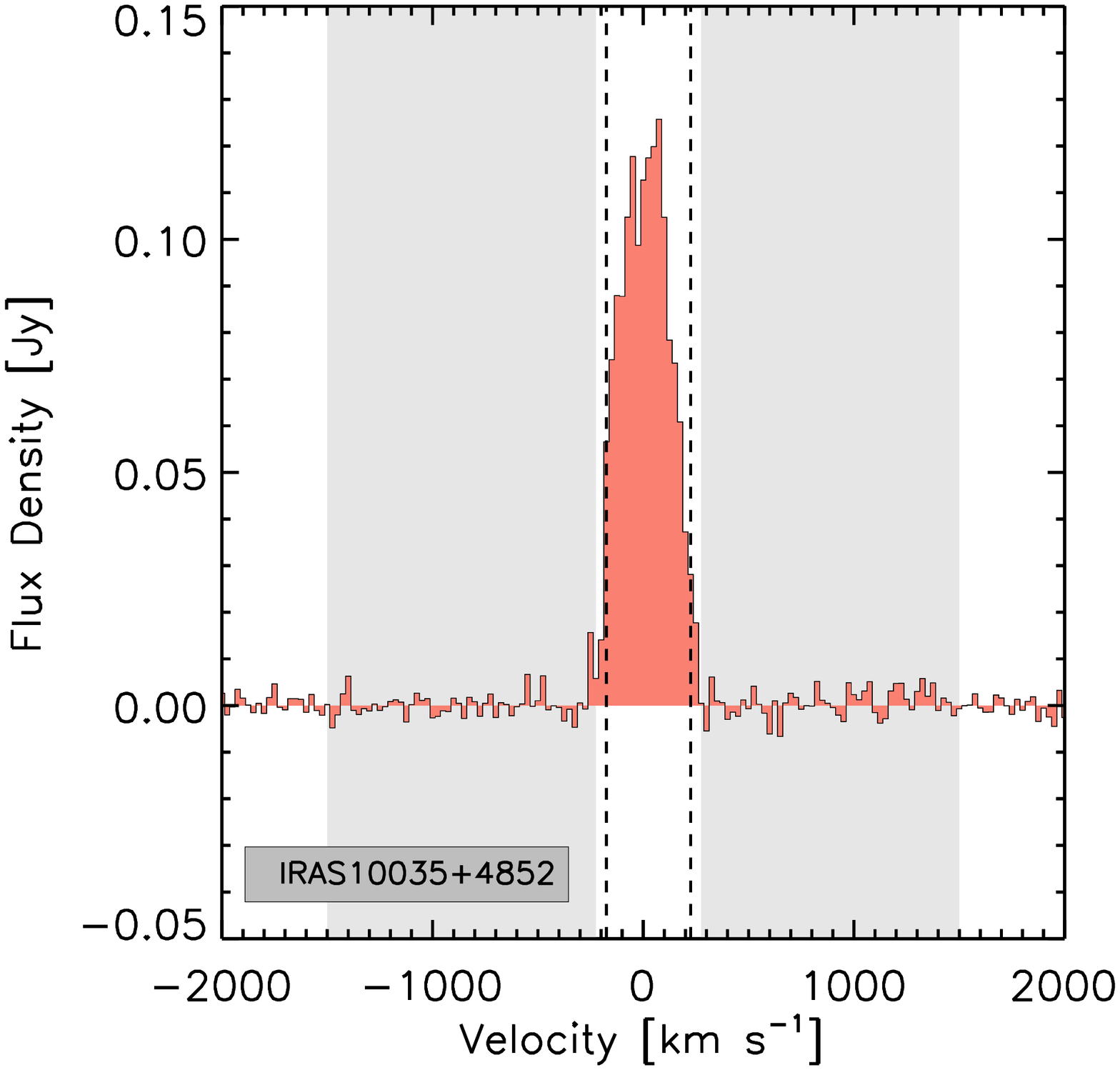}{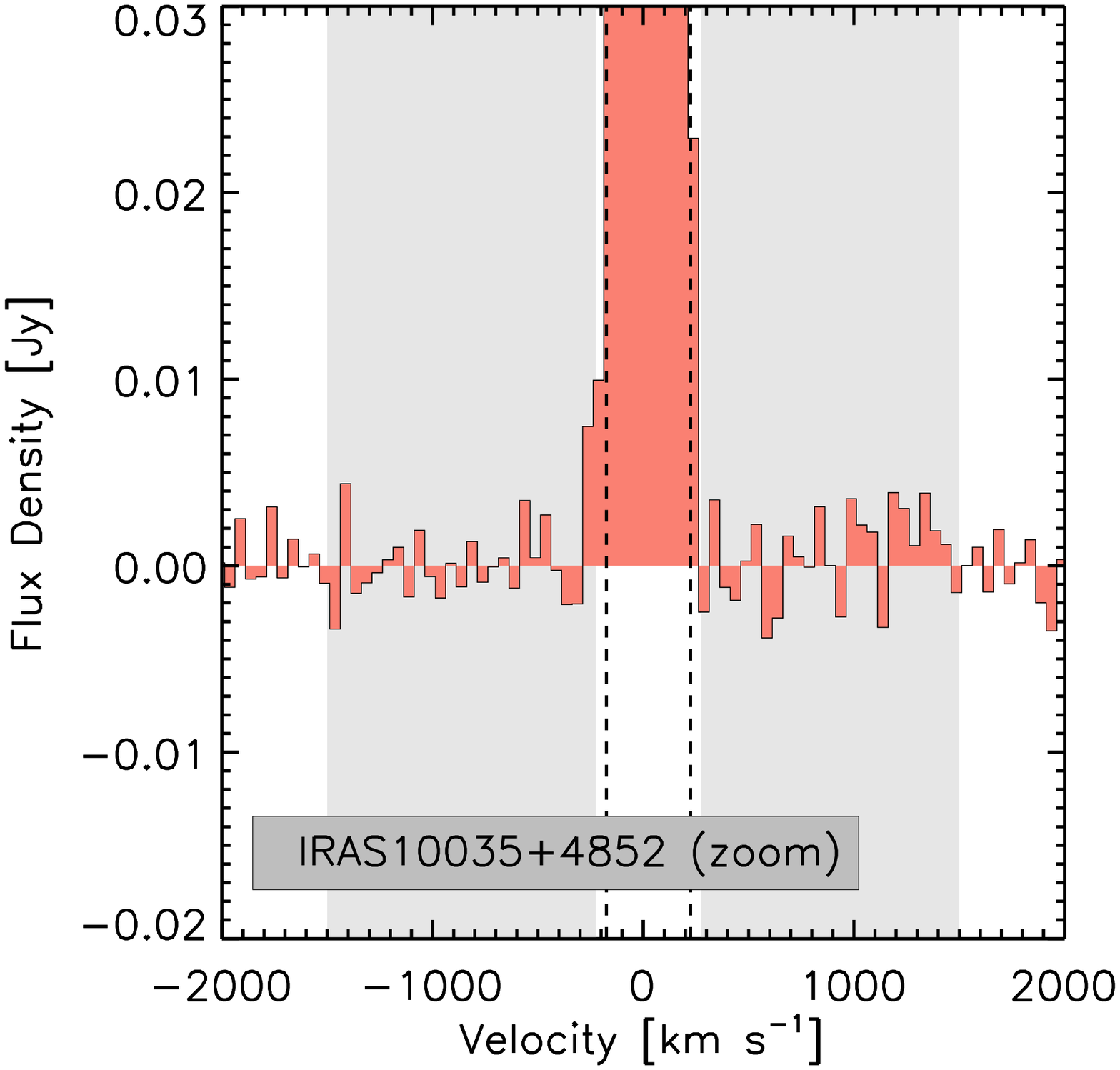}
\caption{Spectra of our sources integrated over the regions of interest shown in Figure \ref{fig:maps}. Each spectrum is shown twice: first showing the full CO line ({\em left}, using 25~km~s$^{-1}$ channels) and then zooming in to focus on faint emission near the level of the noise ({\em right}, using 50~km~s$^{-1}$ channels). The gray region shows the part of the spectrum used to measure the flux in line wings. Dashed vertical lines show $w_{20}$, the width of the profile at 20\% of its peak value. Table \ref{tab:obs} reports integrated quantities for each spectrum.
\label{fig:spectra}}
\end{figure*}

\setcounter{figure}{1}
\begin{figure*}
\plottwo{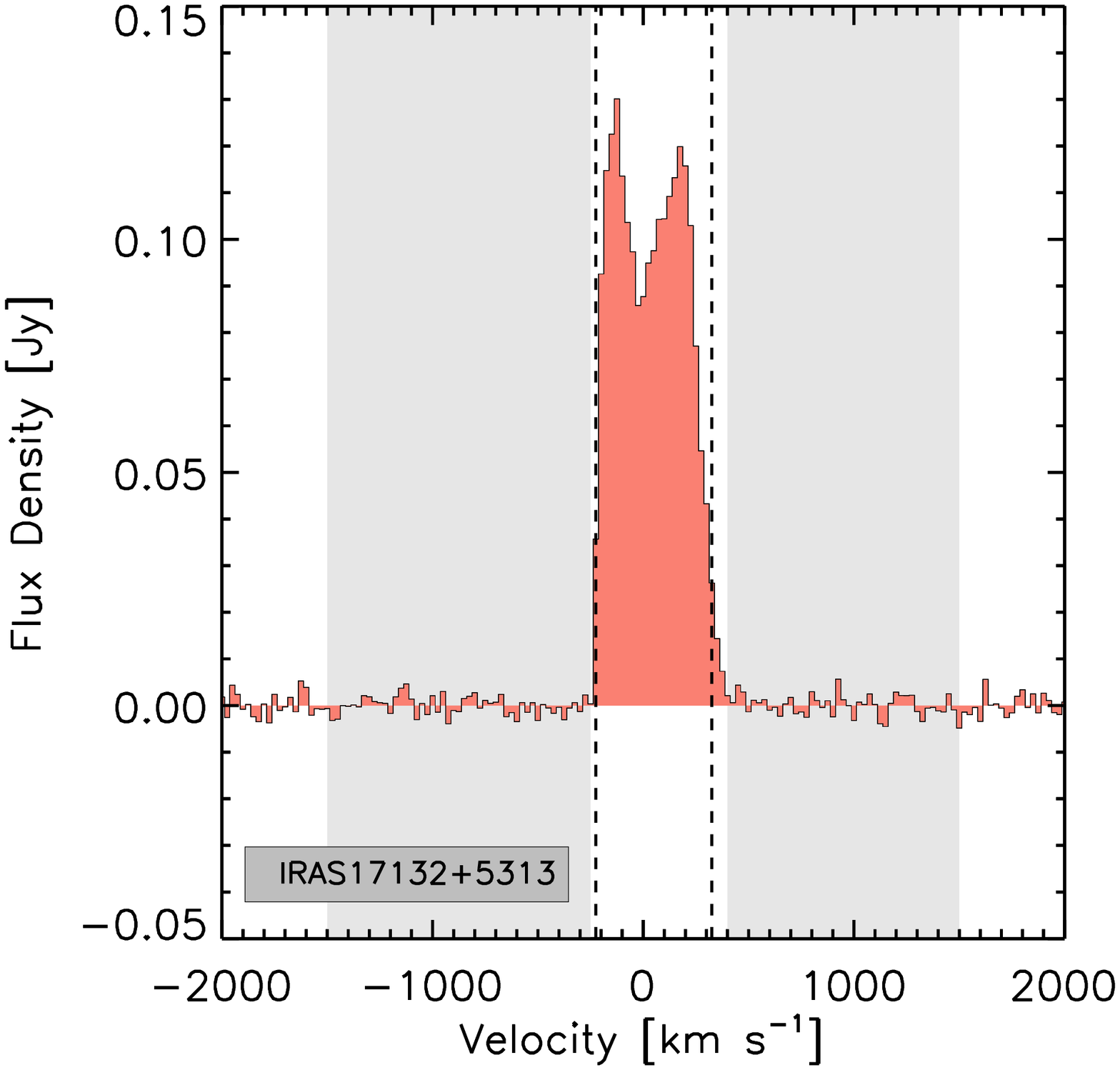}{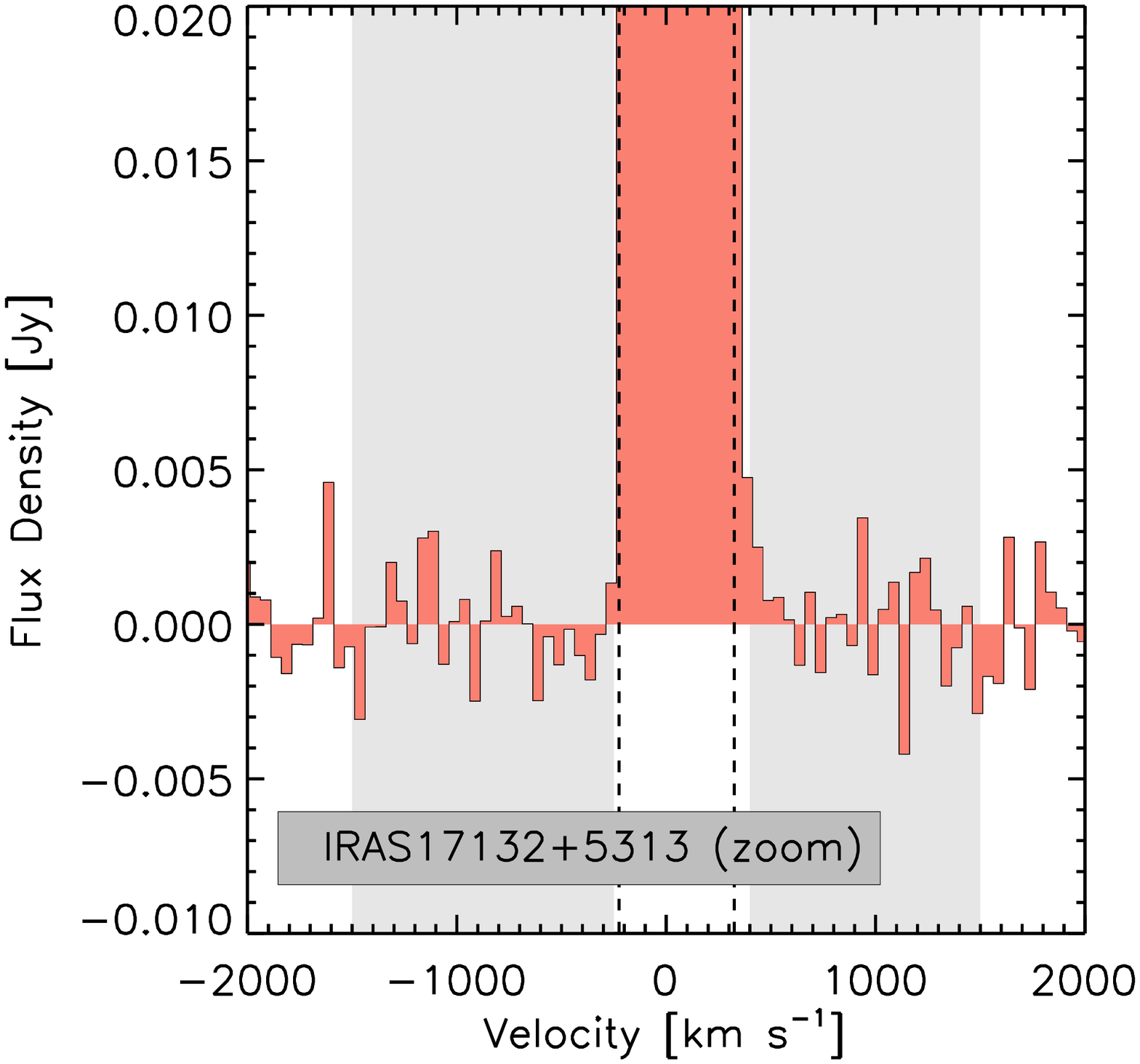}
\plottwo{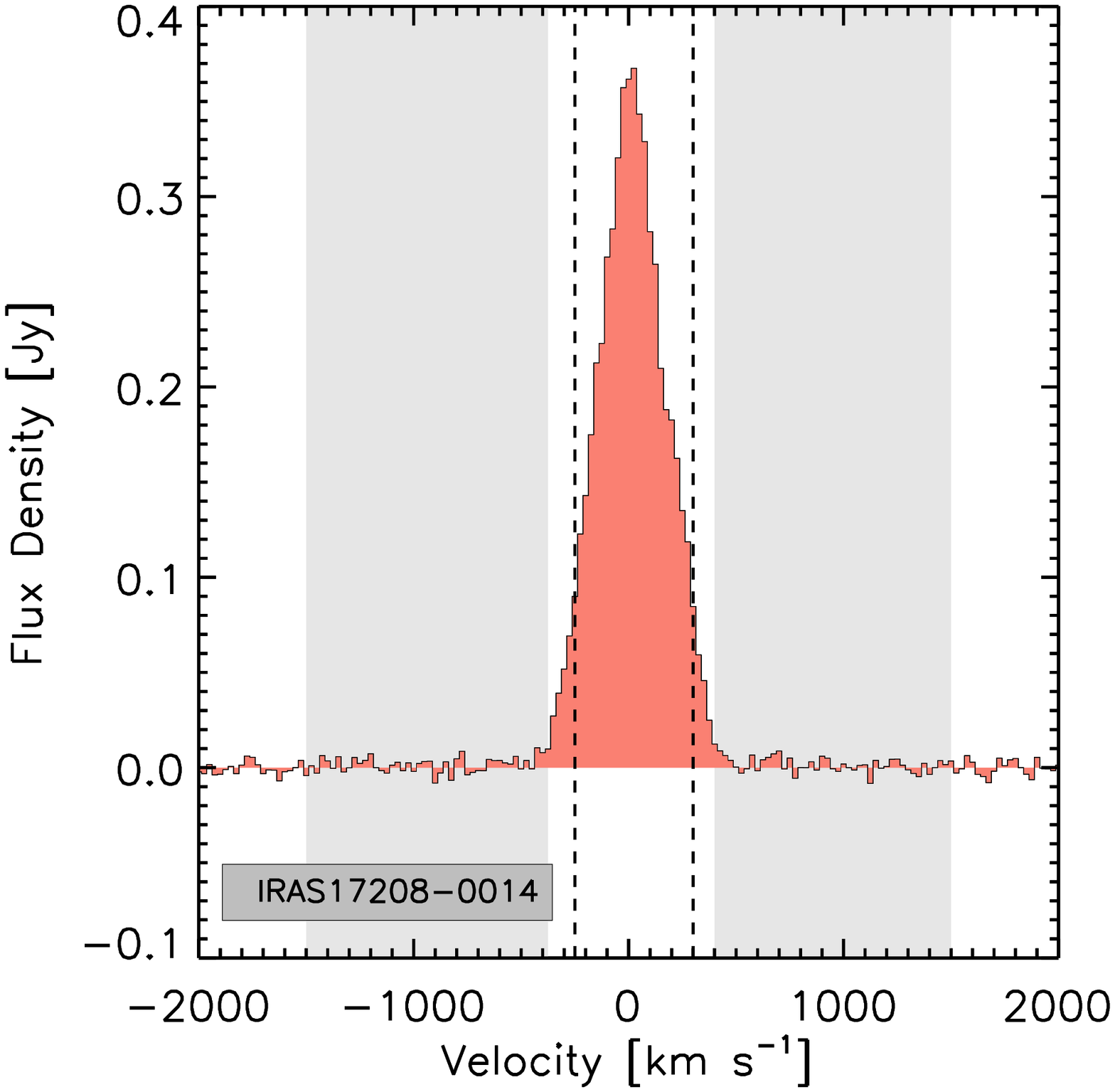}{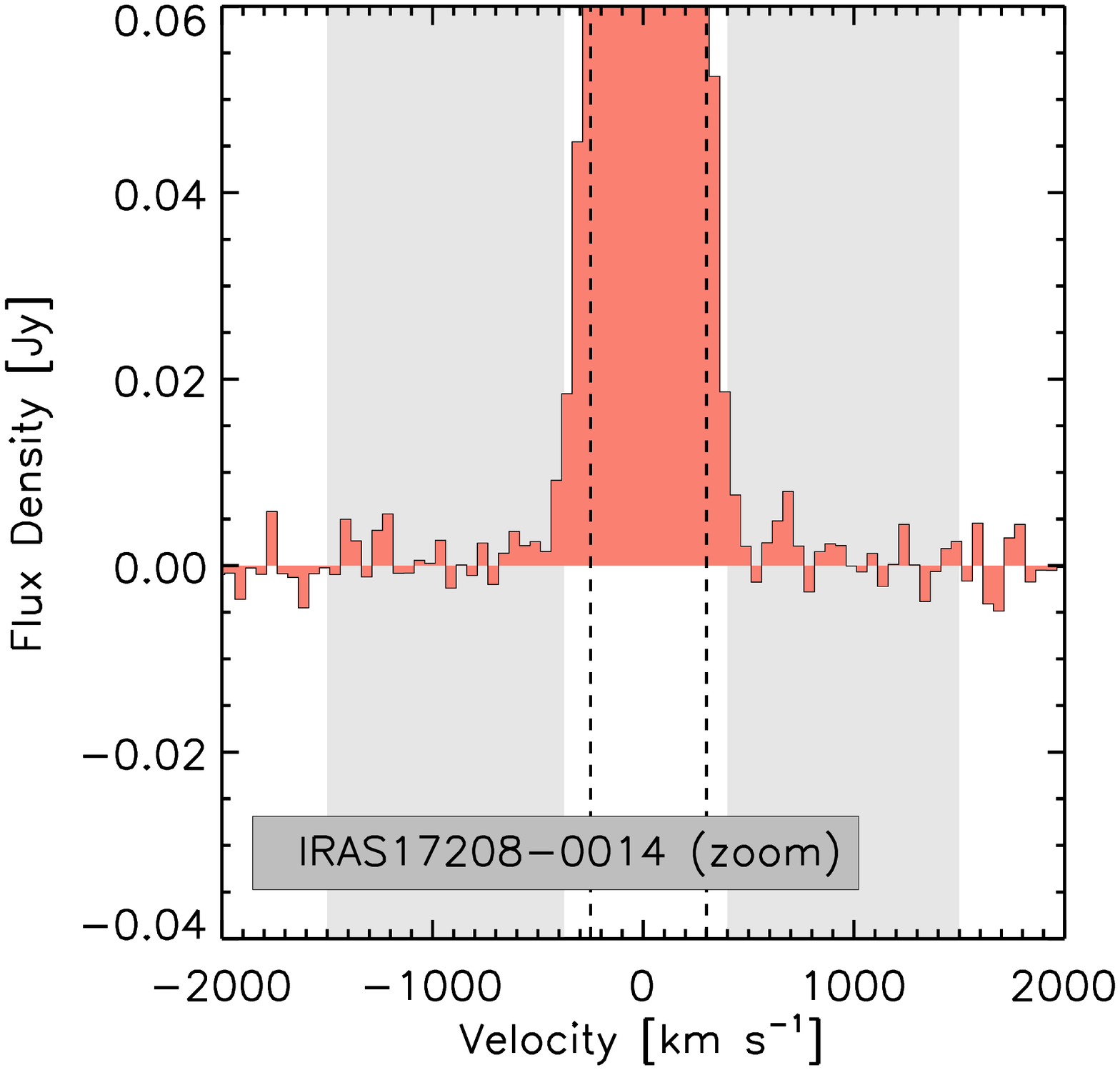}
\caption{(continued).}
\end{figure*}

\begin{figure*}
\plottwo{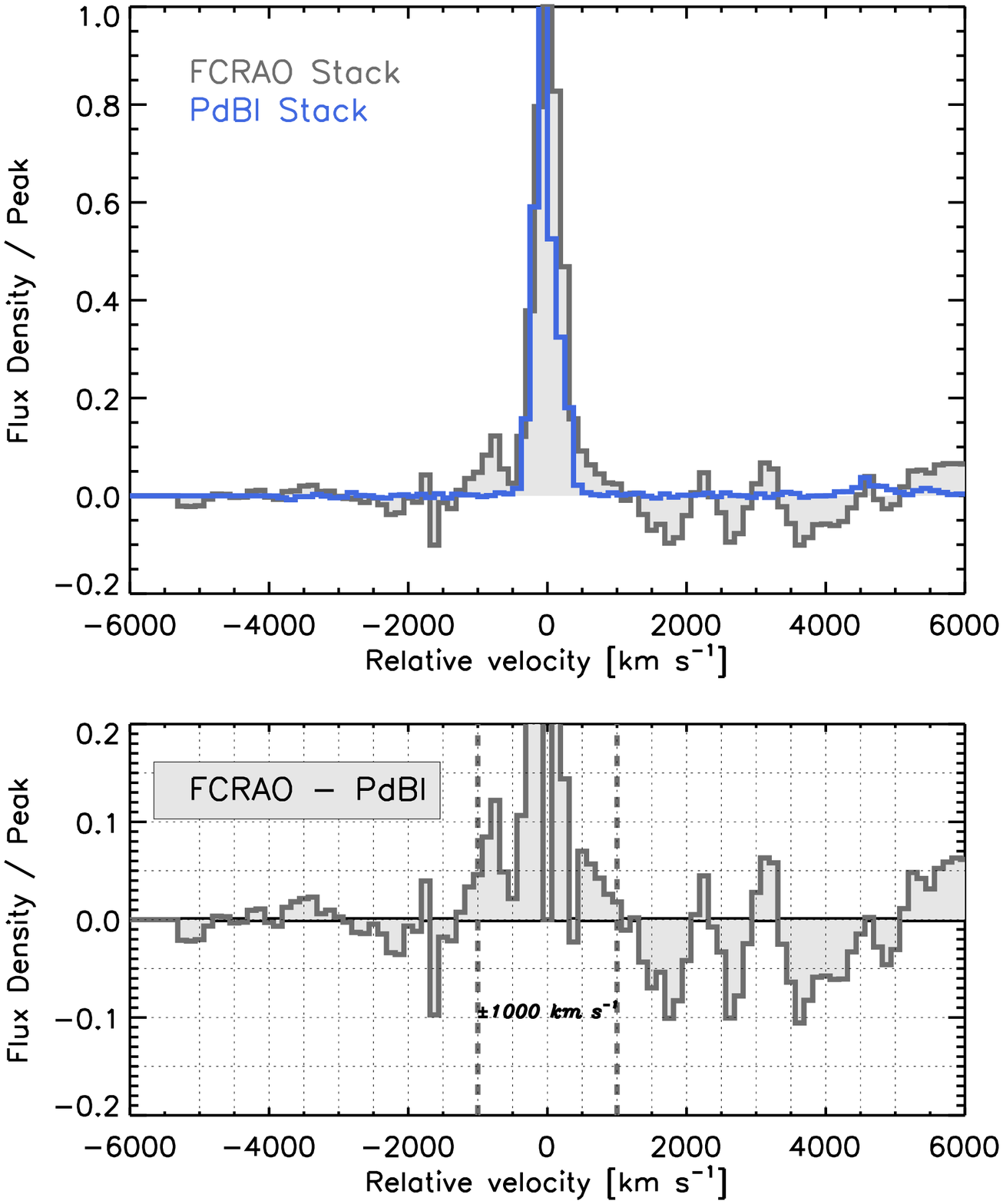}{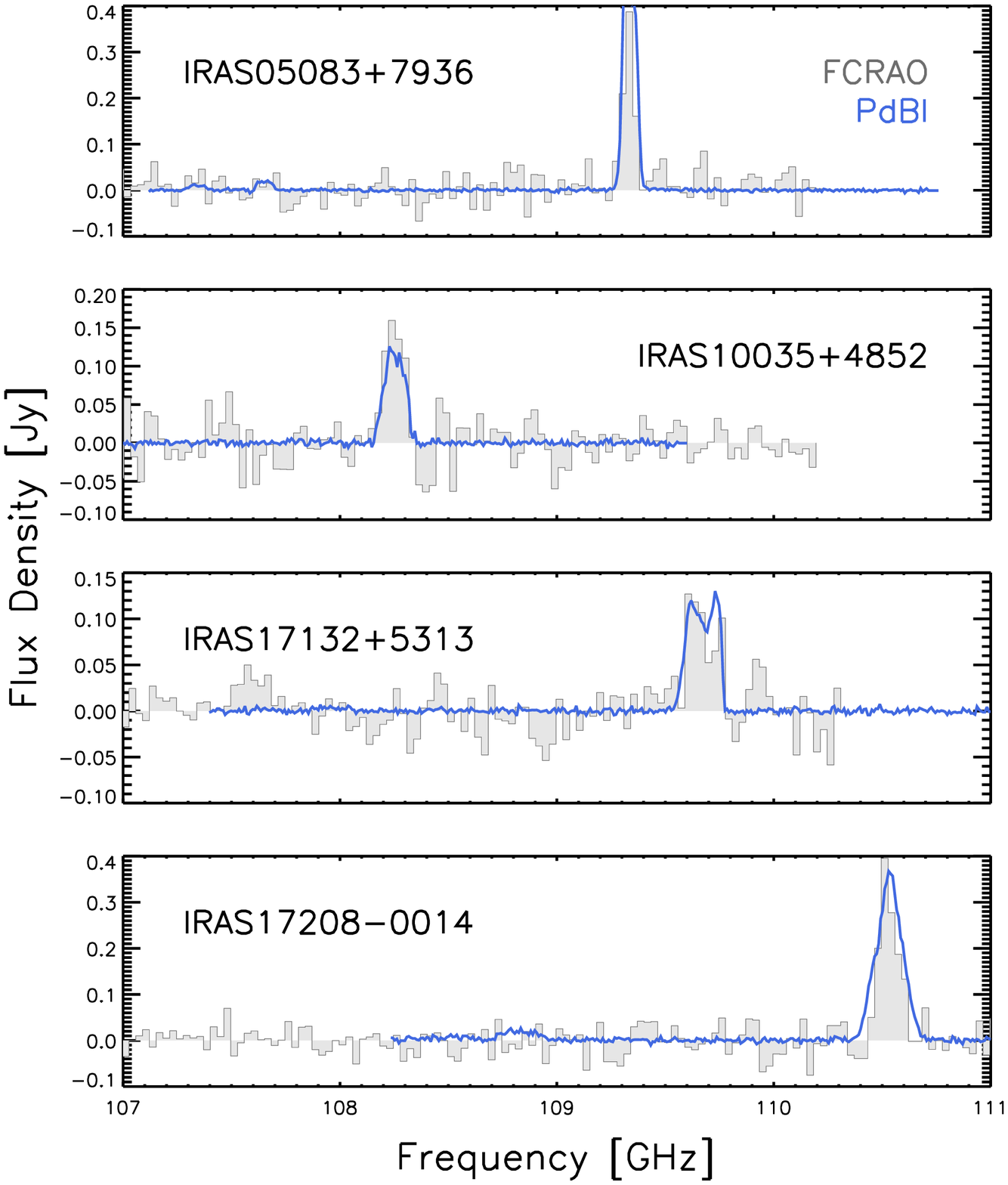}
\caption{
\label{fig:stacks} ({\em left}) Stacked spectra, normalized to the peak of each spectrum from our PdBI observations (blue) over stacked FCRAO spectra of the same four targets from C11 (gray) at 150~km~s$^{-1}$ spectral resolution. As discussed in C11, the stacked FCRAO spectrum exhibits broad line wings at $\approx 10\%$ of the peak flux density. This is evident even in the FCRAO stack of only our four targets. However, the PdBI spectra of the same galaxies, which have an order of magnitude better sensitivity, do not detect these wings or corresponding features in individual spectra (blue) despite a very good match of the observed (main) lines otherwise. The difference between the stacked spectra from FCRAO and the PdBI is shown in the bottom panel. This highlights the presence of a faint, broad component at $\sim 10\%$ of the peak intensity in the FCRAO spectra but not the PdBI spectra. ({\em right}) Comparisons of individual spectra. The PdBI observations match the FCRAO results for the bright part of the line, but the faint line wings implied by the FCRAO stack are absent.}
\end{figure*}

The goal of our observations was to search for the presence of faint line wings. Such wings are not immediately apparent from the images and position-velocity diagrams in Figure \ref{fig:maps}. To search more rigorously, we collapse each cube into a single spectrum, integrating over the region of interest indicated in Figure \ref{fig:maps}. We show these spectra in Figure \ref{fig:spectra} and report their properties in Table \ref{tab:obs}. For each spectrum, we plot two views: a version showing the full line at 25~km~s$^{-1}$ resolution and a version at 50~km~s$^{-1}$ resolution zoomed in to a $y$-axis range $\approx 10$ times the rms noise in the spectrum. The velocity scale in each spectrum is relative to the nominal redshift of the source (the slight offset of IRAS05083 from zero indicates that our adopted redshift for the observations was slight off from the true redshift of the galaxy).

Each source shows broad, bright lines but our observations do not suggest the presence of the wide, faint component seen by C11. In their stacked spectrum, the line wings have average magnitude $\approx 10\%$ of the peak of the spectrum and contribute $\approx 25\%$ of the total CO flux. In each of our individual targets, a wide component 10\% at 10\% of the peak would be several sigma in each $25$~km~s$^{-1}$ channel and extend across $\sim 20$ channels. Such a signal is clearly not present in our data. If the line wings claimed by C11 are real, then they may be present in the other 10 targets that contribute to their stack of star forming galaxies. However, we can confidently reject the signal in four of the brightest members of this sample.

Columns 6 and 7 of Table \ref{tab:obs} quantify the flux in faint line wings in our targets. We report the integrated flux in the spectrum that is within $\pm 1500$~km~s$^{-1}$ of the nominal redshift of the source but outside the velocity range where the main line is $>5\%$ of its peak amplitude. This region appears as the gray area in Figure \ref{fig:spectra}. We report the flux in this velocity range (column 6) and express this as fraction of the total line flux (column 7). Only a few percent of the flux in our targets emerges from faint line wings, in stark contrast to the results found for a larger sample by C11.  In Section \ref{sec:disc}, we discuss this flux in line wings in terms of H$_2$ mass and mass outflow rate.

\subsection{Stacking Comparison}

We explore the contrast with C11 more directly in Figure \ref{fig:stacks}. The right panel shows a direct comparison between the spectra from C11 and the PdBI. For the bright part of the line in individual sources, the two telescopes agree quite well in both amplitude and line shape. In the left panel, we plot the stacked PdBI spectra for our four targets. We follow a procedure similar to C11: we first smooth the spectrum to a resolution of 125~km~s$^{-1}$. Then we identify the peak of the spectrum. We renormalize the peak amplitude of spectrum to unity and recenter the velocity axis so that the peak lies at zero velocity. We grid the spectra onto a grid centered at zero with channels 125~km~s$^{-1}$ wide. Then we coadd all four spectra to produce the averages. We carry out the same procedure for both the PdBI spectra and the FCRAO spectra. We plot both stacks in the left panel of Figure \ref{fig:stacks} and show the difference of the two stacks in the bottom panel. Note that the asymmetric frequency coverage around the line in the FCRAO spectrum leads to variable noise across the velocity range.

The averaged FCRAO spectrum for our four targets shows a broad ($\sim 1000$~km~s$^{-1}$) component with intensity about 10\% of the peak value. This closely resembles the stacked signal presented for 14 galaxies by C11. That is, the stacked FCRAO spectrum for only our four targets does appear consistent with the claim of C11 for their entire star-forming sample.

By contrast, the stack of the PdBI spectra do not show such a feature, so that the residual FCRAO minus PdBI plot in the bottom left panel of Figure \ref{fig:stacks} shows a signal that resembles the C11 claim near zero velocity. Rigorously, we can only say that our observations disagree with the \citet{CHUNG09} spectra regarding faint CO wings for these four targets. We cannot rule out the presence of very bright, broad line wings in the other 10 targets. However given the disagreement for our four present targets we believe the current observations cast some doubt on the broader C11 claim. One possible explanation could be that baseline issues with the single dish telescope (which affect interferometers less) created a spurious signal in the single dish stack and that bad luck lead this signal to not appear in C11's carefully constructed control samples.

\subsection{Faint Broad Wings}

Though we do not find a high-flux extended component, in both I05083 and I17208 we do detect line wings containing at most a few percent of the flux of the main line at marginal significance. These are visible in Figure \ref{fig:spectra} and in the integrated line wing fluxes presented in Table \ref{tab:obs}. Using higher spatial resolution observations, IRAS17208 has recently been demonstrated to have CO outflow \citep{GARCIABURILLO15} and the magnitude and velocity of the outflowing CO emission in their data ($\sim -500$-$-750$~km~s$^{-1}$ and a few mJy) is consistent with the faint emission in our spectrum). To our knowledge, I05083 (VII Zw 31), does not have a previous measurement of faint line wings. Higher resolution observations will be needed to localize the high velocity emission relative to the rotation curve and test whether the wings indeed reflect a CO outflow.

\section{Discussion}
\label{sec:disc}

\begin{figure*}
\plottwo{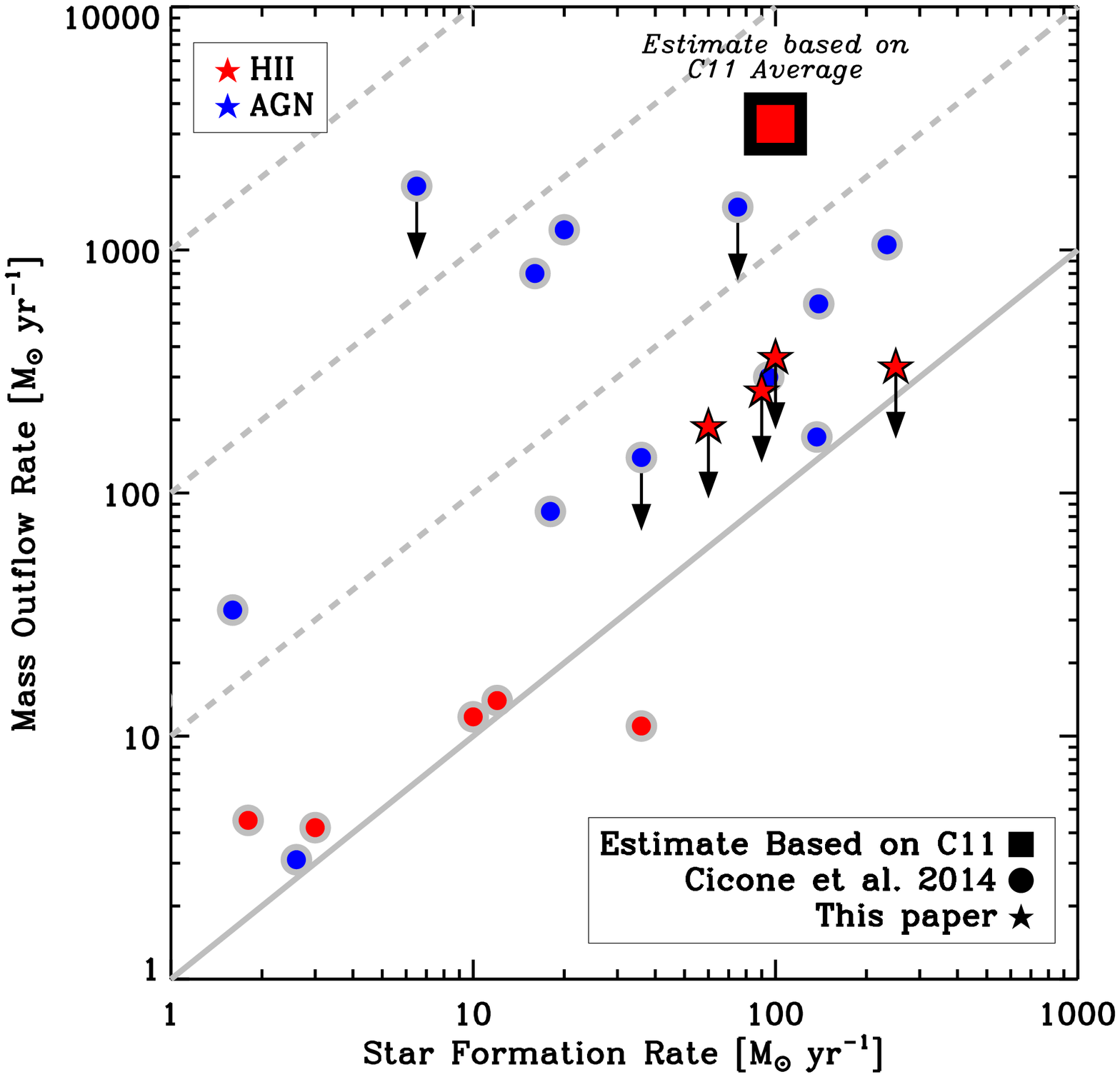}{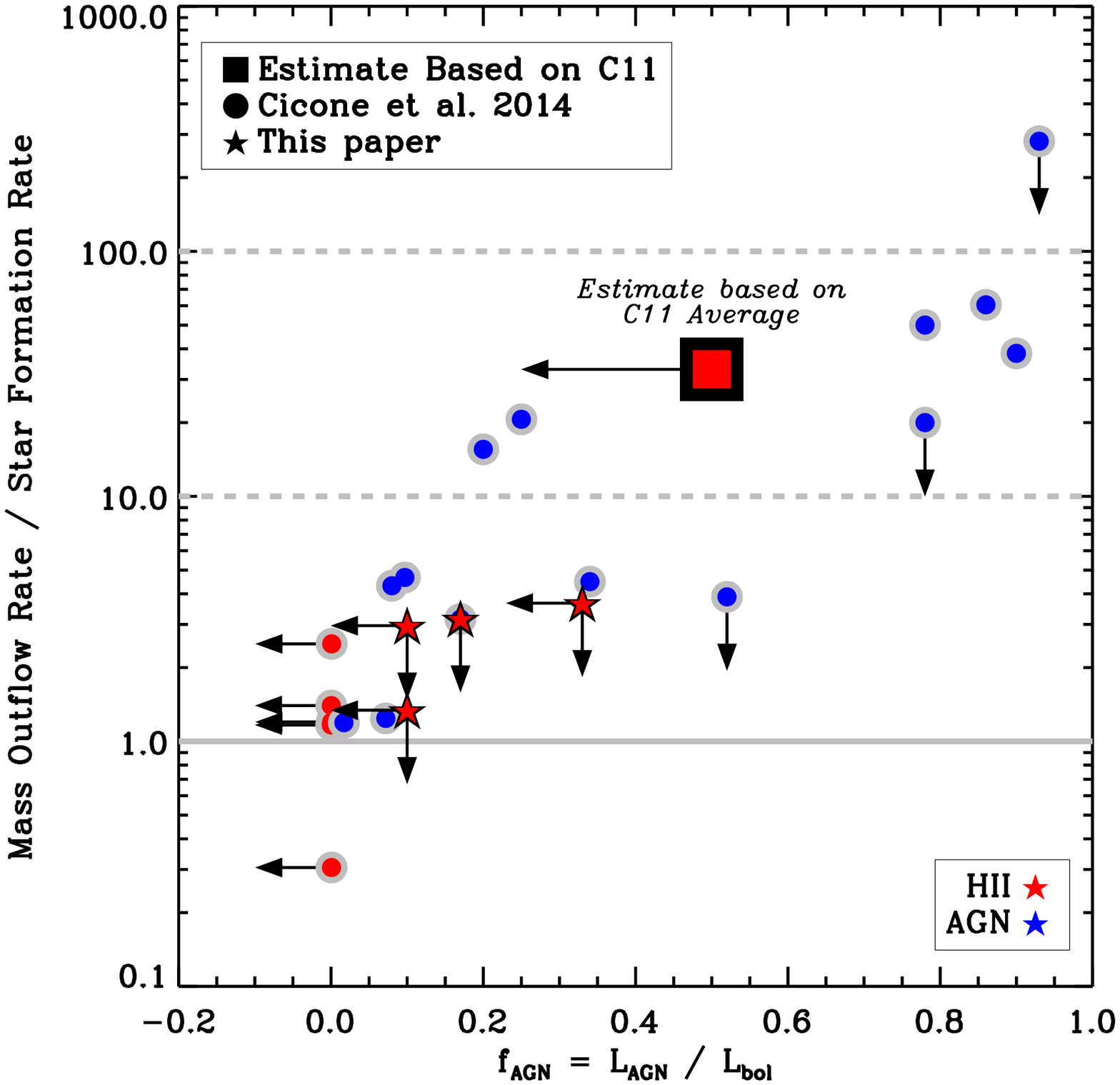}
\caption{
\label{fig:mdot} Molecular outflows, star formation, and AGN activity following the \citet{CICONE14} formalism for $\dot{M}_{\rm OF}$ (Equation \ref{eq:mdot}). ({\em left}) Mass outflow rate as a function of star formation rate. A solid line marks equality, a mass loading factor of unity. Dashed lines show ratios of 10, 100, and 1000. Our new points appear as red circles with black outlines. The \citet{CICONE14} extended sample appears as gray circles marked in blue or red depending on their spectroscopic classification as ``{\sc Hii}'' or ``AGN.'' Our estimate of the average properties implied by the claimed C11 stack appears as a large red square. ({\em right}) The mass loading factor, i.e., the mass outflow rate per unit SFR, as a function of AGN contribution to the total bolometric luminosity. Points and lines are the same as in the left panel. All of our measurements suggest mass loading factors $< 10$. None of the ``{\sc Hii}'' targets in \citet{CICONE14} show a high mass loading factor either. By contrast, reasonable assumptions regarding the C11 stack would imply an average mass loading factor of $\approx 30$, something found only in AGN-dominated systems in \citet{CICONE14}. The observations present here argue against such a high average mass loading factor in star-formation dominated galaxies.}
\end{figure*}

We fail to detect the line wings claimed by C11, which account for $\approx 25\%$ of the total line flux. Instead, we constrain the flux in faint line wings in our targets to be at most a few percent. How do these results, and the claim of C11, compare with the results of the molecular outflow population analysis in \citet{CICONE14}? Qualitatively, that paper argued that high mass loading factors (defined as $\dot{M_{\rm OF}} / SFR$, see below) correlate to high contributions of AGN to the bolometric luminosity, $f_{\rm AGN} = L_{\rm AGN} / L_{\rm bol}$. In order to compare the various results, we adopt the formalism and data compilation from \citet{CICONE14}. They assume a continuous outflow and calculate the mass outflow rate via

\begin{equation}
\label{eq:mdot}
\dot{M}_{\rm OF} = 3 v \frac{M_{\rm OF}}{R_{\rm OF}}
\end{equation}

\noindent with $v$ the mean velocity of outflowing material, $M_{\rm OF}$ the mass associated with the outflow, and $R_{\rm OF}$ the extent of the outflow. As discussed by \citet{CICONE14}, this geometry is likely to yield a high estimate for $\dot{M}_{\rm OF}$ and we suggest that the reader bear this potential bias in mind. The advantage of this approach is that by adopting it, we can make quantitative comparative statements between our measurements and the \citet{CICONE14} ``extended sample.''

We do not have an outflow extent for any of our targets or the C11 sample and so adopt the median $R_{\rm OF} = 600$~pc from the extended sample of \citet{CICONE14}. This is similar to the $R_{\rm OF} = 500$~pc that they use for upper limits. For our targets, we adopt $v \approx 400$~km~s$^{-1}$, approximately the $1\sigma$ velocity for a 1000~km~s$^{-1}$ line width. The same rms velocity is a good description for the faint wings in I05083. For the C11 stack, we adopt $\left<v\right> \approx 500$~km~s$^{-1}$, half of the full extent of their lines, an a reasonable average velocity.

We convert from luminosity to $M_{\rm OF}$ using a \citet{DOWNES98} conversion factor of $\alpha_{\rm CO} = 0.8$~\acounits, again largely adopted for consistency. This implies $M_{\rm OF} < 1$--$3\times 10^8$~M$_\odot$ for our targets. We treat the measured fluxes for I05083 and I17208 as upper limits on $M_{\rm OF}$, reflecting the uncertainty in identifying them as more than outflow candidates based on lower resolution spectra. Using this approach our targets have $M_{\rm OF} \lesssim 1$--$3\times 10^8$~M$_\odot$, which is similar to the typical molecular mass for one of the \citet{CICONE14} outflows. For the C11 targets we take the average luminosity of the starburst subsample and the stacked result of 25\% of the flux in the broad line. This yields an average $M_{\rm OF} \approx 1.5 \times 10^9$~M$_\odot$.

We calculate star formation rates from IR luminosity using the same formula as \citet{CICONE14} for consistency. For the C11 stack, we adopt $\left<SFR\right> \approx 100$~M$_\odot$~yr$^{-1}$, corresponding to a typical total IR luminosity of $10^{12}$~L$_{\odot}$. Table \ref{tab:source} gives AGN fractions or limits for our targets, though we note that these are highly uncertain. We will discuss C11's starburst sample in terms of an upper limit of $f_{\rm AGN} < 0.5$.

Following this approach, our sources have upper limits $ \lesssim 150$--$400$~M$_\odot$~yr$^{-1}$, which is $\lesssim 1$--$4$ times the star formation rate. For the C11 sample, the implied average outflow rate is $\sim 3000$~M$_\odot$~yr$^{-1}$ in the \citet{CICONE14} formalism, which is $\approx 30$ times the SFR of an average individual target. 

As emphasized, the absolute values of these numbers is very approximate. Comparative measurements offer more insight and to this end we plot the results of these calculations along with the \citet{CICONE14} extended sample in Figure \ref{fig:mdot}. The left panel compares the mass outflow rate to star formation rate, so that diagonal lines correspond to mass loading factors ($\dot{M}_{\rm OF}$/SFR). The right panel compares the mass loading factor to the fraction of luminosity in the system supplied by an AGN. In both panels, blue points show galaxies spectroscopically classified as an AGN and red points show starburst (``{\sc Hii}'') systems. The three samples appear as different symbols.

Considering only our points and \citet{CICONE14} (i.e., leaving aside C11), the two plots reinforce the conclusion that mass loading factors above $\sim 10$ are only achieved in systems with a strong AGN contribution. The exact mass loading factors, especially for the star forming systems, may be slightly higher than the values shown --- for consistency all points here assume the same $\alpha_{\rm CO} = 0.8$~\acounits . The geometry may also vary, with poor resolution and inclination effects potentially hiding some outflows. Nonetheless, the data suggest that star formation appears to produce mass loading factors of up to a few, rather than $\sim 10$--$100$. Though we note the lack of a large set of high star formation rate, SFR-dominated targets. By contrast, the $\dot{M}_{\rm OF}$ suggested on average by C11 imply a mass loading factor of $\sim 30$ for their stack of starburst galaxies. This would be as high as the strongest AGN in the \citet{CICONE14} extended sample.

\section{Summary}
\label{sec:conclusions}

The calculations leading to Figure \ref{fig:mdot} require a large number of assumptions, but the figures still summarize the results of this paper. We follow up a claim of extended, high-flux line wings in starburst galaxies by C11. If present, these line wings would imply high mass loading factors in star-formation dominated system, as high as those seen in extreme AGN-dominated systems. We fail to find such features in 4 of the 14 galaxies presented by C11. Instead we limit the flux in such an extended spectral component to less than a few percent. We show that the stacked C11 signal is present in their data for our targets but not in the $\sim 10$ times more sensitive PdBI observations. Two of our targets do show marginally significant flux in an extended line wing: I17208 is known to have an outflow \citep[likely driven by a hidden AGN][]{GARCIABURILLO15}. We identify the other, I05083 (VII Zw 31), as a candidate molecular outflow. Combining these results with the \citet{CICONE14} extended sample, our observations of high SFR, star-formation dominated systems reinforce the idea that high mass loading factors tend to arise in systems where AGN contribute a large fraction of the luminosity. In addition to follow-up on our candidate outflow, sensitive observations of a large set of star-formation dominated high SFR systems will be useful to further test this conclusion.

\acknowledgements {\bf Acknowledgments:} We are indebted to Aeree Chung and Min Yun for generously sharing their FCRAO spectra with us and the anonymous referee for a thorough, helpful report. We acknowledge useful discussions with Paul Martini and the use of the NASA/IPAC Extragalactic Database, NASA's Astrophysics Data Abstract System, and the IDL Astronomy User's Library. Our use of the PdBI relies on the hard work of the IRAM staff, which we gratefully acknowledge.

\appendix

\section{Serenditious CN Measurements}

The wide bandwidth that allows us to search for CO line wings also allows us to detect the CN 1-0 $J=3/2$-$1/2$,$F=5/2$-$3/2$ and $J=1/2$-$1/2$, $F=3/2$-$3/2$ transitions in each target (rest frequencies $113.491$ and $113.191$~GHz). CN is a tracer of dense gas excited in PDR regions; the line strength of the  stronger 113.5~GHz line is typically within a factor of a few of HCN 1-0  \citep[e.g.,][]{AALTO02}. Table \ref{tab:cn} reports the strength of the CN lines for each component and their ratios to one another and CO. The lines are at least tentatively detected for each spectrum. The ratio of the brighter line to the CO line varies between $10$ and $30$, consistent with previouswork. For the high signal-to-noise spectra and the two CN lines show ratios $1.4$--$2.2$. The CN lines appear to have similar width to the CO, implying that the dense gas in these targets may be cospatial with the bulk of the molecular gas traced by CO.

\begin{deluxetable}{lcccc}
\tablecaption{CN 1-0 Observations} 
\tablehead{ 
\colhead{Galaxy} & 
\colhead{113.5 GHz} & 
\colhead{113.2 GHz} & 
\colhead{$\frac{{\rm CN 113.5}}{{\rm CN 113.2}}$} &
\colhead{$\frac{{\rm CN 113.5}}{{\rm CO 1-0}}$} \\
\colhead{} & 
\colhead{(Jy km s$^{-1}$)} & 
\colhead{(Jy km s$^{-1}$)} & 
\colhead{} &
\colhead{}
}
\startdata
I05083 & $5.1 \pm 0.4$ & $3.6 \pm 0.4$ & $1.4$ & $20$ \\
I10035 & \ldots & \ldots & \ldots \\
\ldots comp. A & $0.63 \pm 0.24$ & $0.53 \pm 0.24$ & \ldots & $\approx 30$ \\
\ldots comp. B & $0.87 \pm 0.32$ & $0.47 \pm 0.32$ & \ldots & $\approx 25$ \\
I17132 & \ldots & \ldots & \ldots \\
\ldots comp. A & $1.6 \pm 0.3$ & $0.86 \pm 0.33$ & $1.9$ & $30$ \\
\ldots comp. B & $0.58 \pm 0.13$ & $0.26 \pm 0.13$ & \ldots & $\approx 10$ \\
I17208 & $11.2 \pm 0.8$ & $5.0 \pm 0.8$ & $2.2$ & $13$ \\
\enddata
\label{tab:cn}
\end{deluxetable}

\end{document}